%% file: main.tex
\documentclass[aps,prl,reprint,superscriptaddress]{revtex4-2}

\usepackage{amsmath}
\usepackage{amsfonts}
\usepackage{amssymb}
\usepackage{graphicx}
\usepackage{booktabs}
\usepackage{bibunits}
\defaultbibliographystyle{apsrev4-2}

\begin{document}

\title{Quantum Metric Connects Bond Delocalization to Phonon Softening}

\author{Tianqi Deng}
\email{dengtq@zju.edu.cn}
\affiliation{State Key Laboratory of Silicon and Advanced Semiconductor Materials, School of Materials Science and Engineering, Zhejiang University, Hangzhou 310058, China}

\author{Tiejun Zhu}
\affiliation{State Key Laboratory of Silicon and Advanced Semiconductor Materials, School of Materials Science and Engineering, Zhejiang University, Hangzhou 310058, China}

\date{\today}

\begin{abstract}
Delocalized bonding electrons have long been associated with soft bonds and low phonon frequencies. However, a quantitative connection between electron delocalization and phonon softening remains elusive. We identify such a connection through the momentum-space quantum metric, which measures the intrinsic delocalization of the occupied electronic states. We show that a quantum geometric contribution to phonon softening is proportional to the integrated quantum metric. First-principles calculations of more than a hundred compounds demonstrate that greater quantum metric generally accompanies softer phonons. Localized Wannier orbitals reveal the bonding environments associated with this delocalization. In ferroelectric GeTe, LiBeSb, and BaTiO$_3$, polar distortion localizes the bonding orbitals, reduces quantum metric, and stabilizes the soft mode of the nonpolar structure. In PbSe, pressure-induced topological band inversion induces an enhanced quantum metric and phonon softening without change in crystal symmetry and bonding. These results establish a common quantum geometric interpretation of electron delocalization and phonon softening across distinct bonding environments.
\end{abstract}

\maketitle

\paragraph{Introduction}
Phonon softening connects the microscopic behavior of chemical bonds to thermal transport, dielectric response, ferroelectricity, and structural instability~\cite{Han2023PhononSoftening,Riffe2025OpticalPhononScattering,Cao2024OpticalPhononSoftening,Cohen1992Ferroelectricity,Matsunaga2011VibrationalSoftening,Shen2023PrecursorPhononSoftening}. Different pictures of chemical bonding have been invoked to explain soft modes in materials with widely different chemistries~\cite{Neaton2005BiFeO3,Lee2014ResonantBonding,Wuttig2024MetavalentHypervalent,Hempelmann2021LongRangeForces,Cao2024LonePairAnharmonicity}. A recurring idea is that localized electrons are associated with stiff bonds, while more delocalized bonding electrons can reduce the restoring force on displaced atoms~\cite{Feynman1939Forces}. Quantifying this idea microscopically requires a measure of electronic delocalization that also enters the interatomic forces.

Quantum geometry provides a natural language for this problem. Integrated over the Brillouin zone, the momentum-space quantum metric gives the gauge-invariant part of the Wannier spread and quantifies the intrinsic delocalization of the occupied electronic states~\cite{Souza2000Localization,MarzariVanderbilt1997,Marzari2012MLWF,Yu2025QuantumGeometry}. Recent studies have identified geometric contributions to electron--phonon coupling and force constants~\cite{Yu2024EPCGeometry,Pellitteri2025PhononGeometry}, as well as a mixed momentum--displacement geometry underlying strain-induced ferroelectricity~\cite{Hu2025EPCGeometry}. However, a direct quantitative connection between electronic delocalization and phonon softening has not been established.

In this work, we identify a quantum geometric connection between bond delocalization and phonon softening. We show that the quantum metric of the occupied states enters the interatomic force constants (IFCs) through the electronic response to atomic displacements. First-principles calculations show that greater electronic delocalization generally accompanies softer phonons in diverse crystal systems. Wannier bond orbitals reveal the chemical bonding associated with this delocalization and a common picture of ferroelectric softening in GeTe, LiBeSb, and BaTiO$_3$. Pressure-induced band inversion in PbSe extends the connection to a topological transition, where an enhanced quantum metric accompanies phonon softening relative to the compression trend.

\paragraph{Quantum metric and phonon softening}
A rigid-ion crystal, where electrons follow the ions without internal redistribution, sets the classical reference for lattice stiffness. The classical Born--Land\'e model illustrates how ionic charge and coordination affect transverse optical (TO) phonons~\cite{SupplementalMaterial}, thus explaining the correlation between ionicity and phonon softness~\cite{Ioffe1964Ionicity,Loh1968OpticalPhonons,Spitzer1970ThermalConductivity}. Electronic redistribution induced by atomic displacement supplies an additional response beyond this classical reference. This response reshapes the short-range bonding interaction and produces anomalous Born effective charges for long-range interaction, both of which can contribute to phonon softening~\cite{Cohen1992Ferroelectricity,Zhong1994GiantLOTO,Cao2024OpticalPhononSoftening}.

To assess how electronic redistribution affects the force constants, we compare the crystal density-response contribution $\Phi^{\mathrm{resp}}$ with its rigid-ion counterpart at the same structure. The crystal contribution is~\cite{SupplementalMaterial}
\begin{equation}
 \Phi^{\mathrm{resp}}_{\kappa\alpha,\kappa'\beta}
 =2g_s\sum_{n\in\mathrm{occ}}\int[d\mathbf{k}]\,
 \operatorname{Re}\langle u_{n\mathbf{k}}^{(\tau_{\kappa\alpha})}|
 \partial_{\kappa'\beta}v^{\mathrm{ion}}|u_{n\mathbf{k}}\rangle .
 \label{eq:main:ifc-response}
\end{equation}
Here $\kappa$ labels atoms, $\alpha,\beta$ are Cartesian directions, $u_{n\mathbf{k}}$ is a cell-periodic Bloch state, and $u^{(\tau_{\kappa\alpha})}$ is its displacement-induced tangent projected outside the occupied manifold. The spin multiplicity is $g_s$, and the reciprocal volume element is $[d\mathbf{k}]=V_{\mathrm{cell}}d^3k/(2\pi)^3$. Here $v^{\mathrm{ion}}$ is the ionic potential energy of an electron. In the rigid-ion counterpart, each reference density moves rigidly with its ion, as described in the Supplemental Material (SM)~\cite{SupplementalMaterial}.

To connect this response to localization, recall the integrated occupied-manifold quantum metric $\mathcal G$~\cite{Yu2025QuantumGeometry,Jiang2025NonlinearQuantumGeometry}:
\begin{align}
\mathcal G^{\alpha\beta}
 &=\operatorname{Re}\sum_{\substack{n\in\mathrm{occ}\\m\notin\mathrm{occ}}}
 \int[d\mathbf{k}]\,Q_{mn\mathbf{k}}^{\alpha\beta} .
 \label{eq:main:metric}\\
  Q_{mn\mathbf{k}}^{\alpha\beta}
 &=\langle\partial_{k_\alpha}u_{n\mathbf{k}}|u_{m\mathbf{k}}\rangle
 \langle u_{m\mathbf{k}}|\partial_{k_\beta}u_{n\mathbf{k}}\rangle,
 \label{eq:main:transitionqgt}
\end{align}
The tensor $\mathcal G$ is invariant under changes of basis within the occupied manifold. Its trace gives $\operatorname{Tr}\mathcal G=\Omega_I$, where $\Omega_I$ is the gauge-invariant Wannier spread per cell~\cite{MarzariVanderbilt1997,Marzari2012MLWF}. Therefore, for gapped insulators, semiconductors, and semimetals where valence and conduction bands do not cross, the trace of $\mathcal G$ can be directly evaluated using Wannier functions of the occupied bands. For metals which are not discussed in this work, the quantum metric may also be evaluated directly from wave functions. A larger integrated metric therefore indicates greater intrinsic delocalization of the occupied bonding electrons.

\begin{figure}[tbp]
\includegraphics{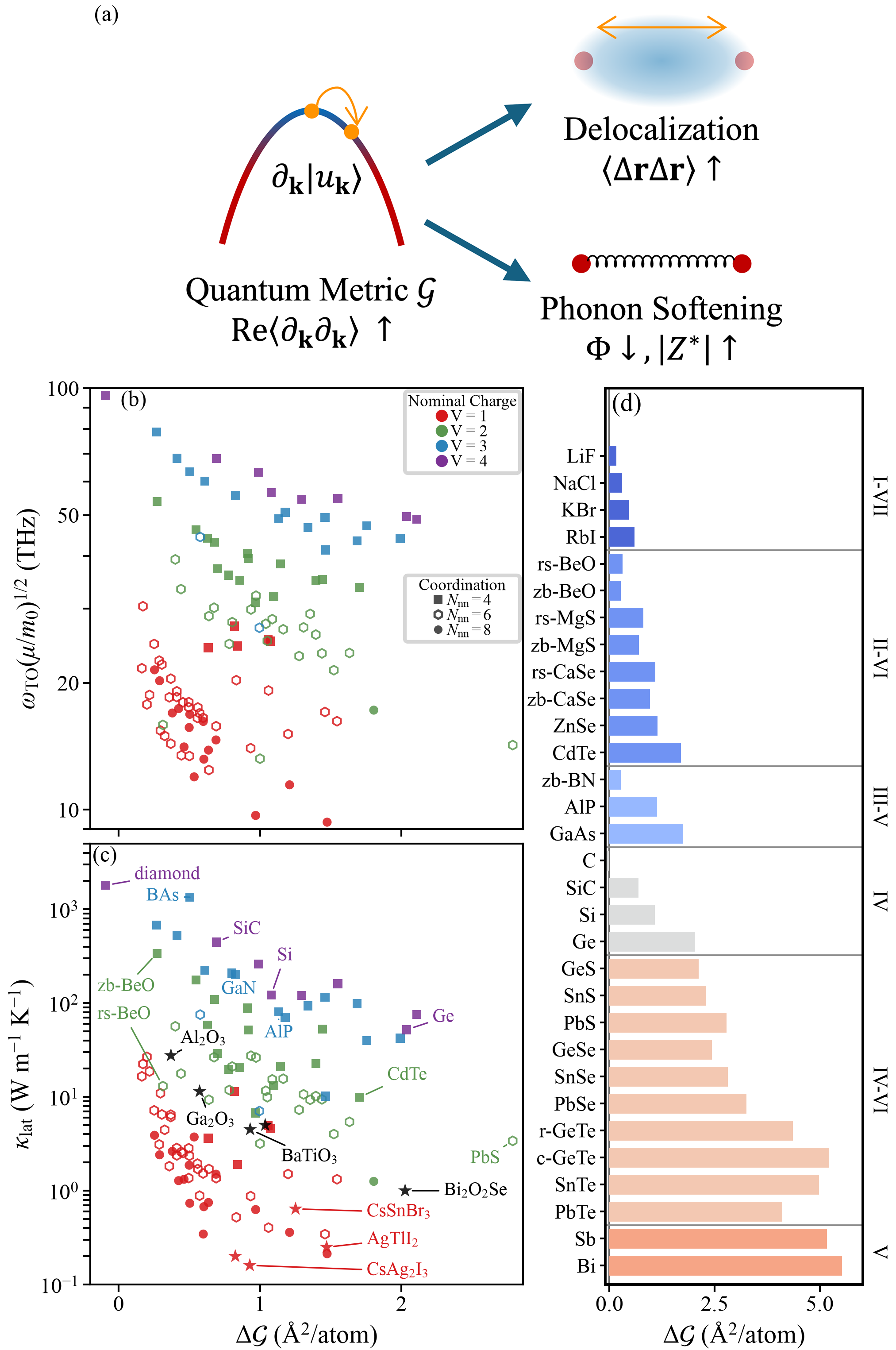}
\caption{\label{fig_bond}(a) Illustration of quantum geometric nature of delocalization, reduced stiffness, and anomalous BECs. (b) Mass-scaled TO frequency and (c) room-temperature lattice thermal conductivity versus $\Delta\mathcal G$. Colors denote valency group $V$ and symbols denote the coordination number $N_{\mathrm{nn}}$. Stars in (c) mark additional complex halides (red) and oxides (black). (d) Chemical and structural trends in the signed $\Delta\mathcal G$ of representative compounds.}
\end{figure}

We connect $\Phi^{\mathrm{resp}}$ to the quantum metric by extending the leading-order matrix-element approximation of Hu \textit{et al.}~\cite{Hu2025EPCGeometry} to both ionic and screened potential perturbations, giving
\begin{align}
 \Phi^{\mathrm{resp}}_{\kappa\alpha,\kappa'\beta}
 &\simeq-2g_s\sum_{\substack{n\in\mathrm{occ}\\m\notin\mathrm{occ}}}
 \int[d\mathbf{k}]\,
 \frac{\tilde\gamma_{mn,\mathbf{k}}^{\kappa\alpha}
 \gamma_{mn,\mathbf{k}}^{\kappa'\beta}}{\Delta_{mn\mathbf{k}}}
 \operatorname{Re}Q_{mn\mathbf{k}}^{\alpha\beta}
 \nonumber\\
 &\simeq-C_{\kappa\alpha,\kappa'\beta}\mathcal G^{\alpha\beta} .
 \label{eq:main:ifc-qgt}
\end{align}
The transition energy is $\Delta_{mn\mathbf{k}}=E_{m\mathbf{k}}-E_{n\mathbf{k}}>0$, while $\gamma$ and $\tilde\gamma$ are the ionic and screened force-gradient factors~\cite{Hu2025EPCGeometry}. Replacing them by effective values gives $C_{\kappa\alpha,\kappa'\beta}=2g_s\tilde\gamma_{\mathrm{eff}}^{\kappa\alpha}\gamma_{\mathrm{eff}}^{\kappa'\beta}/\Delta_{\mathrm{eff}}$. This central result represents a momentum-space quantum geometric contribution to the electronic response, weighted by coupling strengths and excitation energies. The same quantum metric enters the electronic dielectric response and the electronic contribution to the Born effective charges (BECs)~\cite{SupplementalMaterial}. These responses share the same quantum geometric origin, linking spatial delocalization with both lattice softening and polarizability [Fig.~\ref{fig_bond}(a)].

To illustrate the phonon consequence of Eq.~\eqref{eq:main:ifc-qgt}, consider the relative optical displacement of a one-dimensional alternating cation--anion chain with nearest-neighbor IFC. Electronic IFC corrections $\Delta\Phi_{\mathrm{nn}}$ between nearest neighbors and the acoustic sum rule give a quantum geometric softening contribution proportional to the quantum metric
\begin{equation}
 \left.\Delta\omega_{\mathrm{O}}^2\right|_{\mathrm{geom}}\mu
 = -2\Delta\Phi_{\mathrm{nn}}
 \simeq 2C_{\mathrm{nn}}\Delta \mathcal G^{xx}
 \propto -\Delta \mathcal G^{xx},
 \label{eq:main:mode}
\end{equation}
where $\mu$ is the reduced mass, $\Delta\Phi_{\mathrm{nn}}$ is the electronic response-induced non-rigid IFC contribution, the approximation uses the same effective coefficient \(C_{\mathrm{nn}}\) per bond for the non-rigid and rigid-ion reference, and $\Delta \mathcal G^{xx}$ is the longitudinal quantum metric gain per cell relative to the rigid-ion reference. In an alternating cation--anion chain, the signs of the ionic force gradients are expected to be opposite, $\gamma_+ \gamma_-<0$~\cite{SupplementalMaterial}. If screening preserves the sign, $C_{\mathrm{nn}}<0$. Thus a positive $\Delta \mathcal G^{xx}$ leads to a softening $\left.\Delta\omega_{\mathrm{O}}^2\right|_{\mathrm{geom}}<0$ through Eq.~\eqref{eq:main:mode}. The chain illustrates how delocalization along a bond can soften an optical mode. For the materials comparison, we use the trace of the metric gain per atom as a direction-independent measure of the change in electronic delocalization upon bond formation. The Su–Schrieffer–Heeger (SSH) model further illustrates that in the equivalent-site limit with short-range interaction, the integrated quantum metric diverges as the hopping alternation vanishes and the gap closes, accompanying the electronic instability toward Peierls distortion~\cite{SupplementalMaterial}.

We test the relation between bonding electron delocalization and phonon softening in more than 100 gapped two-atom structures spanning different nominal charges and coordination environments [Fig.~\ref{fig_bond}(b)]. We compare compounds using $\Delta\mathcal G=\left(\operatorname{Tr}\mathcal G-\sum_\kappa\operatorname{Tr}\mathcal G_{\kappa}^{\mathrm{ion}}\right)/N_{\mathrm{ion}}$, where $\mathcal G_{\kappa}^{\mathrm{ion}}$ is the metric of isolated ion $\kappa$ and $N_{\mathrm{ion}}$ is the number of ions per cell. This signed difference measures the change in occupied-manifold spread per atom relative to the isolated-ion reference. A positive value indicates greater collective delocalization in the crystal. We scale the frequency by $\sqrt{\mu/m_0}$, with $m_0$ the atomic mass unit, to eliminate the explicit mass dependence and compare the effective bond stiffness across compounds. Figure~\ref{fig_bond}(b) shows that a larger $\Delta\mathcal G$ generally accompanies a lower mass-scaled TO frequency within each nominal-charge and coordination group, while the groups are separated according to the classical ionic Born--Land\'e baseline~\cite{SupplementalMaterial}. The within-group trend is consistent with the geometric softening picture, linking increased bond delocalization with lower phonon frequencies. Separate plots for each valence and coordination number group are given in the SM~\cite{SupplementalMaterial}.

Softer phonons can also reduce lattice thermal conductivity by lowering phonon velocities and increasing the phase space for scattering. Figure~\ref{fig_bond}(c) shows a corresponding trend across materials, from high-conductivity diamond and BAs to low-conductivity halides. Within the group-IV elemental and binary compounds, conductivity decreases as $\Delta\mathcal G$ increases. The isoelectronic compounds BAs, GaN, and AlP also show different values of $\Delta\mathcal G$ and conductivity, illustrating the diversity of transport behavior among related bonding environments. Comparing zincblende and rocksalt BeO further highlights the combined influence of coordination and delocalization~\cite{Cao2024OpticalPhononSoftening}. The comparison extends to materials with more complex bonding environments. The low-conductivity halides CsAg$_2$I$_3$~\cite{Shen2026Ultralow}, CsSnBr$_3$~\cite{Xie2020Halide}, and AgTlI$_2$~\cite{Zeng2024Thermal} and the oxide Bi$_2$O$_2$Se~\cite{Gibson2021Thermal} lie at relatively high $\Delta\mathcal G$. The group IV--VI and group V thermoelectric materials also have large metric gains [Fig.~\ref{fig_bond}(d)], consistent with a connection between delocalized bonding, soft phonons, and low thermal conductivity~\cite{Cao2024LonePairAnharmonicity,Lee2014ResonantBonding,Wuttig2024MetavalentHypervalent,Hempelmann2021LongRangeForces}. Together, these examples show how the quantum metric connects electronic delocalization to lattice softness across diverse materials, with consequences for thermal transport. Computational procedures and data are given in the SM~\cite{SupplementalMaterial}.

\paragraph{Resolving into chemical bonds}

\begin{figure}[tbp]
 \includegraphics{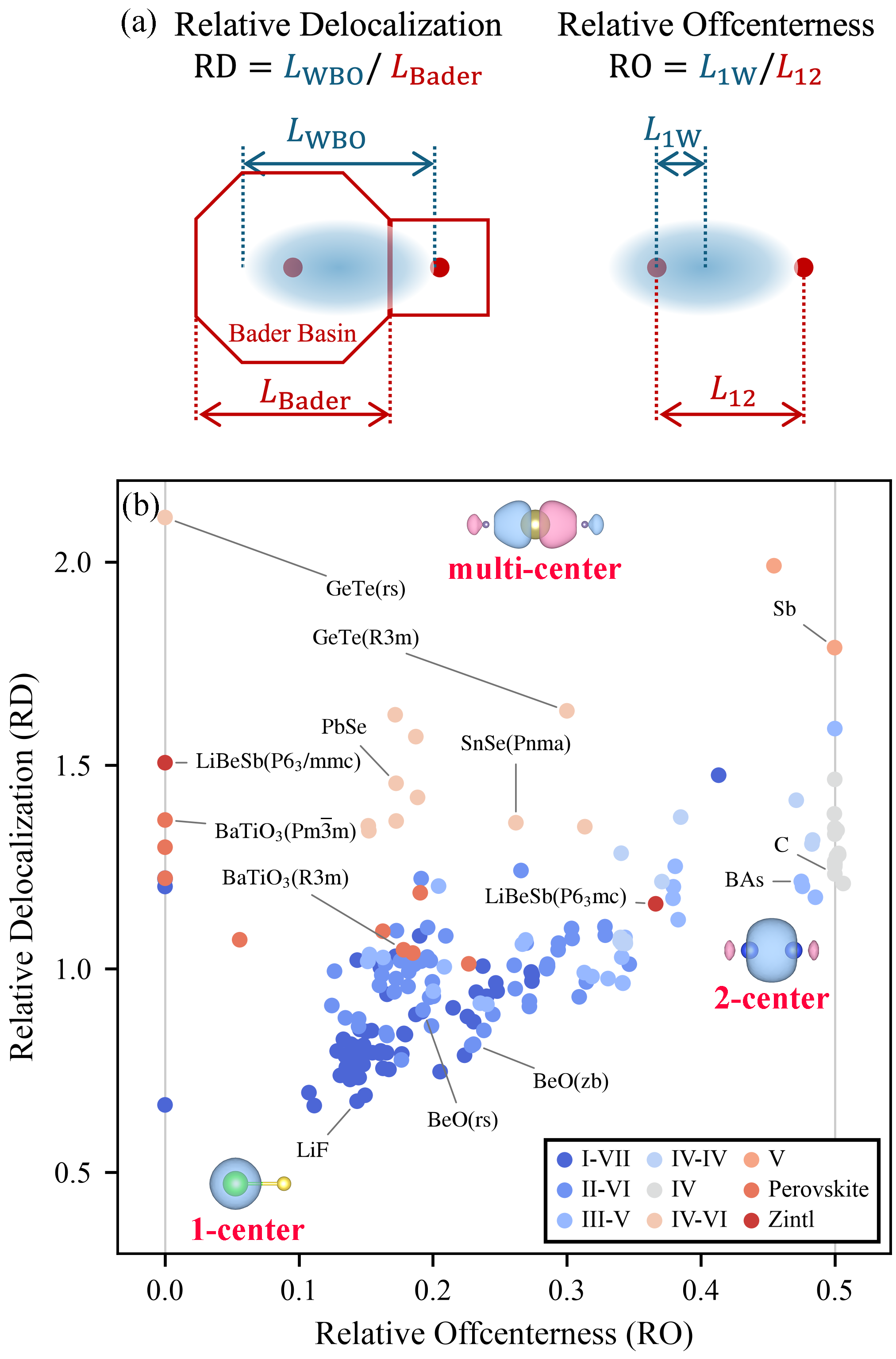}
 \caption{\label{fig_rdro}Wannier bond orbitals (WBOs) reveal the bonding environments associated with delocalization. (a) Schematic definition of Relative Delocalization (RD) and Relative Offcenterness (RO) of WBOs. (b) The RD--RO map of the WBO with highest RD in each material. The map connects one-center states, two-center covalent bonds, and extended multi-center bonds.}
\end{figure}

Localized Wannier orbitals reveal the bonding environments associated with the collective delocalization measured by the quantum metric. We refer to the Wannier orbitals of the occupied bands as \textit{Wannier bond orbitals} (WBOs), which provide a crystalline localized-bond picture analogous to natural bond orbitals in molecules~\cite{Reed1988NBO,Dunnington2012PeriodicNBO,Marzari2012MLWF}. The connection between the orbital picture and the metric follows from the total Wannier spread,
\begin{equation}
 \sum_{i}\Omega_{i}=\Omega_I+\widetilde\Omega
 =\operatorname{Tr}\mathcal G+\widetilde\Omega,
 \label{eq:main:total-spread}
\end{equation}
where $\sum\Omega_{i}$ sums the spreads of all occupied WBOs in a cell, and $\widetilde\Omega\geq0$ is the gauge-dependent remainder~\cite{MarzariVanderbilt1997,Marzari2012MLWF}. Maximally localized Wannier orbitals minimize this remainder while leaving $\Omega_I$ unchanged. The metric therefore measures the intrinsic collective delocalization of the occupied manifold, while the localized orbitals show how the electronic states are distributed among atoms and bonds. Equally extended orbitals can be centered on an atom or shared between neighboring atoms, so their centers and principal extents supply chemical information beyond the scalar trace.

We characterize each WBO using relative delocalization (RD) and relative offcenterness (RO),
\begin{equation}
 \mathrm{RD}=\frac{L_{\mathrm{WBO}}}{L_{\mathrm{Bader}}},\qquad
 \mathrm{RO}=\frac{L_{\mathrm{1W}}}{L_{\mathrm{12}}} .
 \label{eq:main:rdro}
\end{equation}
Here $L_{\mathrm{WBO}}$ measures the orbital's principal spatial extent, $L_{\mathrm{Bader}}$ characterizes the nearest atom's Bader basin size, $L_{\mathrm{1W}}$ is the distance from the WBO center to that atom, and $L_{\mathrm{12}}$ is the separation of the two nearest atoms [Fig.~\ref{fig_rdro}(a)]. Precise definitions and localization procedures are given in the SM~\cite{SupplementalMaterial}. Unlike $\mathcal G$, the individual orbital descriptors depend on the Wannier gauge. Symmetry-adapted Wannier orbitals preserve crystal symmetry, but their spreads depend on the choice of centers and irreducible representations~\cite{Sakuma2013SymmetryAdapted}. For a common gauge convention across materials, we use unconstrained maximally localized Wannier functions~\cite{Marzari2012MLWF}, initialized with selected columns of the density matrix (SCDM)~\cite{Damle2018Disentanglement,Damle2017SCDMk}, unless otherwise stated.

The RD--RO map [Fig.~\ref{fig_rdro}(b)] shows the WBO with the highest RD in each material, highlighting an orbital that is extended relative to its nearest atom. The orbitals form a continuous distribution among one-center, two-center, and multi-center bonding environments. Compact, predominantly one-center orbitals lie toward $\mathrm{RD}<1$ and small $\mathrm{RO}$, as in ionic LiF. Two-center covalent bonds lie near $\mathrm{RO}=1/2$, with RD of order unity, as in diamond and BAs. More extended orbitals lie above and to the left of the ionic--covalent band, as in cubic GeTe, where they reveal multi-center bonding. Polar covalent BN and lone-pair-containing SnSe occupy intermediate regions. The map thus gives a chemical picture of how delocalization is expressed in different bonding environments, complementing the collective measure provided by $\mathcal G$.

\paragraph{Phase Transitions}

\begin{figure}[tbp]
 \includegraphics{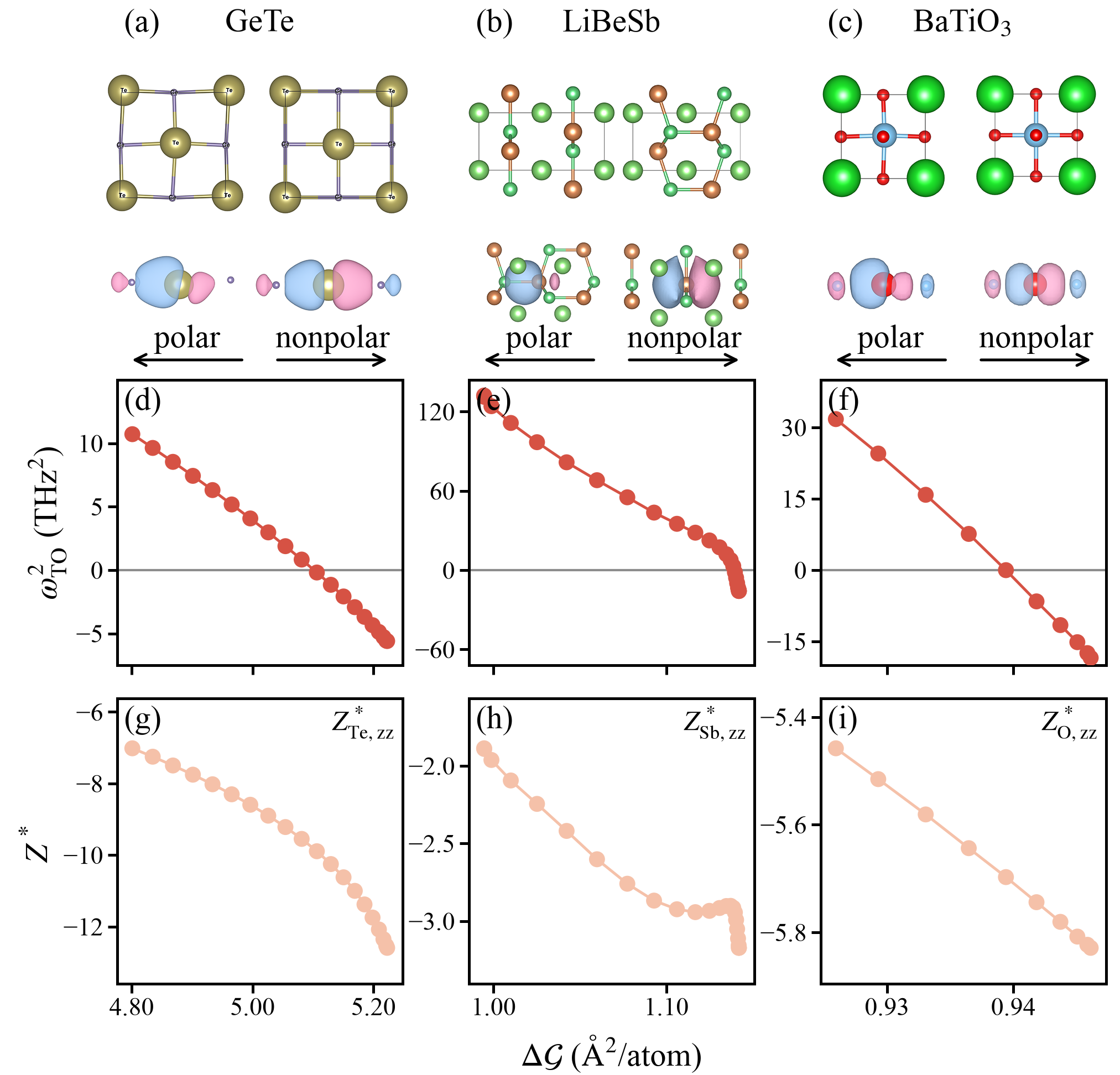}
 \caption{\label{fig_fe}Ferroelectric distortion localizes bonds and stabilizes the soft mode. Structures and WBOs for (a) GeTe, (b) LiBeSb, and (c) BaTiO$_3$ connect the polar and nonpolar limits. From nonpolar to polar (right to left), (d--f) $\omega_{\mathrm{TO}}^2$ rises through zero. (g--i) Signed Born effective charges $Z^*_{zz}$ for Te, Sb, and O, respectively, in units of $e$. Negative $\omega_{\mathrm{TO}}^2$ denotes an imaginary frequency.}
\end{figure}

\begin{figure}
 \includegraphics{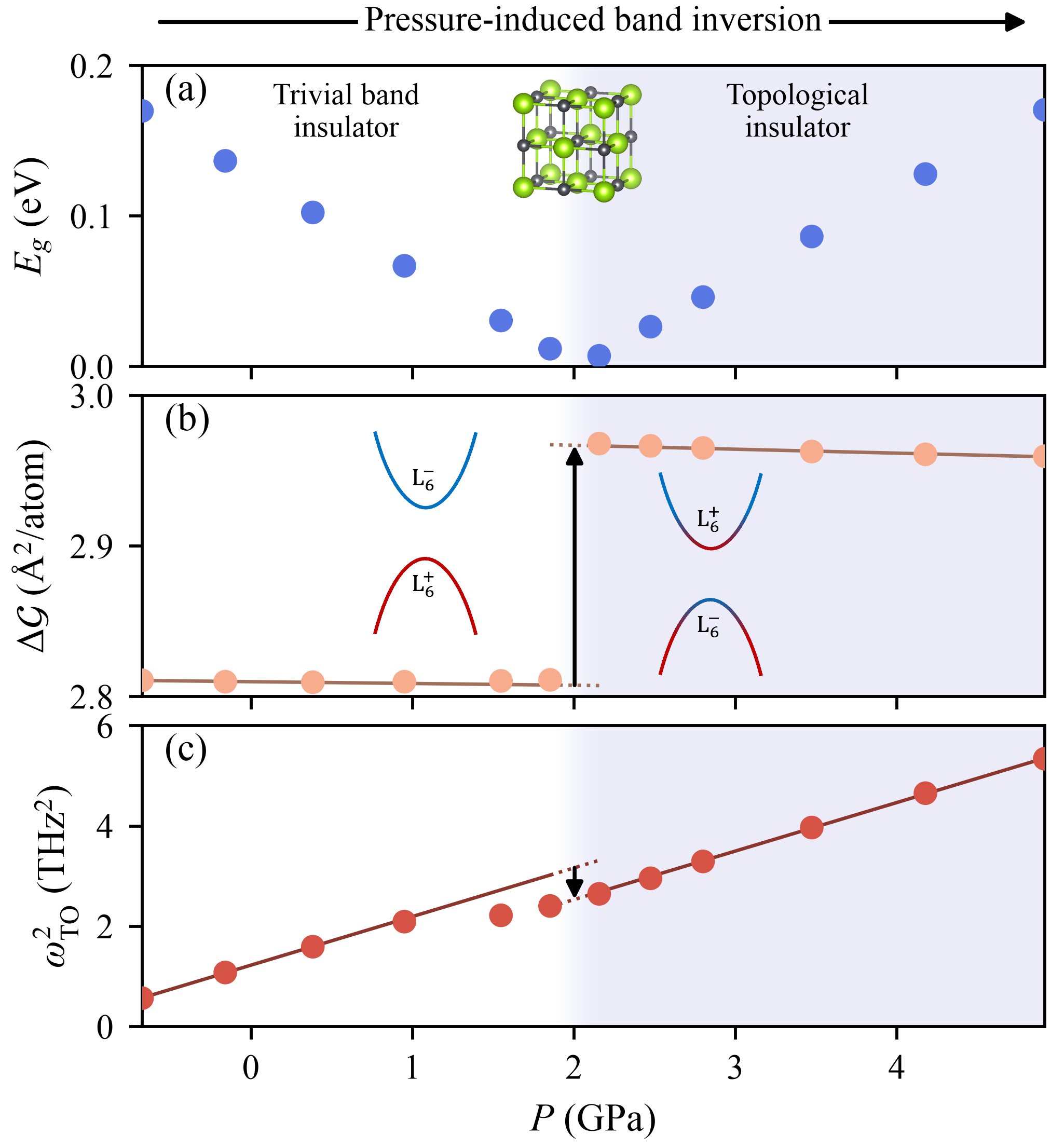}
 \caption{\label{fig_topo}Quantum metric and TO phonons across pressure-induced band inversion in PbSe. (a) Pressure dependence of the direct gap at $L$. (b) $\Delta\mathcal G$, with schematics of the $L_6^+$ and $L_6^-$ band ordering reported by Chen \textit{et al.}~\cite{Chen2019PbSeTopology}. (c) $\omega_{\mathrm{TO}}^2$ versus pressure. The solid lines are linear guides to the eye, the dotted segments are their extrapolations, and the arrows mark the offsets. Shading denotes the inverted phase.}
\end{figure}

While comparisons across chemical families establish a broad connection between electronic delocalization and phonon softness, a structural path within a single material allows us to follow their evolution together. The composition, nominal ionic references, and atomic masses remain fixed while the bonding evolves during a phase transition. Changes in $\Delta\mathcal G$ then track changes in the occupied manifold as the structure distorts. These paths connect the bond map to the microscopic process of a structural instability.

We first consider the ferroelectric distortions of GeTe, LiBeSb, and BaTiO$_3$~\cite{Wdowik2014SoftPhonon,Cohen1992Ferroelectricity,Bennett2012Hexagonal}. In the RD--RO map [Fig.~\ref{fig_rdro}(b)], the polar and nonpolar phases of each material occupy distinct regions. The nonpolar phases have more extended, atom-centered WBOs, while the polar phases lie closer to the localized ionic--covalent band. The orbital images in Fig.~\ref{fig_fe}(a--c) show the corresponding change in bonding. In each material, the representative orbital changes from a more extended, atom-centered form with multi-center character in the nonpolar structure to a more localized, off-center form in the polar structure. Along the same paths, $\Delta\mathcal G$ decreases as the TO mode hardens and $\omega_{\mathrm{TO}}^2$ rises through zero [Fig.~\ref{fig_fe}(d--f)]. The anion BEC magnitudes are also smaller at the polar endpoints [Fig.~\ref{fig_fe}(g--i)], consistent with the reduced electronic polarizability accompanying localization. Together, these results suggest a common mechanism across the three bonding environments: electronic delocalization contributes to the soft mode of the nonpolar phase, while the polar distortion localizes the bonding electrons and stabilizes the mode~\cite{Wuttig2024MetavalentHypervalent, Hempelmann2021LongRangeForces, Bennett2012Hexagonal, Cohen1992Ferroelectricity}.

Band topology offers a complementary perspective on electronic delocalization. In topological band insulators, the occupied states can be obstructed from forming exponentially localized Wannier functions that respect the protecting symmetries, connecting the real-space picture of bonding to the global organization of electronic states in momentum space~\cite{Bradlyn2017TopologicalQuantumChemistry}. This connection raises the question of how a topological transition affects the quantum metric and, in turn, lattice dynamics. In rocksalt PbSe, hydrostatic pressure induces a topological transition through band inversion at $L$, providing a setting in which to examine this question~\cite{Chen2019PbSeTopology}. Near 2~GPa, the calculated direct gap at $L$ narrows to a minimum and then reopens [Fig.~\ref{fig_topo}(a)], accompanied by a sharp increase in $\Delta\mathcal G$ [Fig.~\ref{fig_topo}(b)]. As the gap narrows, mixing between the bands near $L$ can enhance the momentum derivatives entering the quantum metric. Across the transition, the $L_6^+$ and $L_6^-$ band characters exchange. The TO mode hardens with pressure on both sides of the crossover. Across the crossover, however, $\omega_{\mathrm{TO}}^2$ shifts below the extrapolated low-pressure trend [Fig.~\ref{fig_topo}(c)], coinciding with the increase in $\Delta\mathcal G$. This deviation is consistent with an additional geometric softening contribution across the topological transition, even as pressure hardens the mode overall.

\paragraph{Discussion}
We have established a quantum geometric connection between bond delocalization and phonon softening through the occupied-state quantum metric. First-principles calculations validated this connection across diverse material families and along structural and electronic transitions within different individual materials. The integrated metric measures the intrinsic delocalization of the occupied manifold, while Wannier bond orbitals reveal its chemical expression in different bonding environments. Their relative delocalization and offcenterness provide a quantitative map of bonding character. This understanding of the chemical origin confirms delocalization as a common route towards phonon softening.

A promising route towards such delocalization is inspired by rocksalt phases, including BeO and GeTe. For rocksalt BeO, the increased coordination number reduces the ionic baseline TO frequency according to the Born--Land\'e model. The mode is further softened by the increased quantum metric shown in Fig.~\ref{fig_bond}(d). In the case of IV-VI rocksalt structures like GeTe, the multi-center bond is likely enforced by crystal symmetry, similar to the case of a filling-enforced obstructed atomic insulator \cite{Xu2024FillingEnforced}. The rocksalt structure has six equivalent nearest neighbors, which requires 6 equivalent filled WBOs if they are localized 2-center bonds. However, IV-VI compounds only have 10 (5 pairs of) valence electrons, which cannot fill such 6-fold 2-center bonds. Therefore, symmetry and filling enforce occupation of the atom-centered $s$-like lone-pair orbitals on the cation and anion, and 3 anion-centered $p$-like multi-center orbitals. This enforced delocalization may be responsible for the soft phonons across IV-VI and group V systems with rocksalt or distorted rocksalt structures, and could inspire crystal-symmetry and filling-based chemical bond design in other systems.
The PbSe example further connects the effect to topological band inversion, extending the study of delocalization and lattice softness into quantum materials beyond chemical bond pictures. Various design principles of topologically nontrivial materials could also be applied to introduce delocalization-based phonon softening to extend its application beyond quantum materials. It could also play some role in the observed coincidence that many important thermoelectric materials, which favor low thermal conductivity, are topological insulators\cite{Chen2019PbSeTopology,Zhang2009TopologicalInsulators}.  Recent progress in measuring the quantum metric tensor~\cite{Kang2025QuantumGeometry,Kim2025QuantumMetric,CardenasCastillo2024OpticalSum} may provide a route to experimentally probe this connection. As a ground-state property accessible to \textit{ab initio} calculations, the quantum metric offers a common framework for investigating lattice softness in thermal, ferroelectric, and phase-change materials. 

\paragraph{Acknowledgment}
This work was supported by the National Natural Science Foundation of China (Grant No. 62574179) and Zhejiang Provincial Natural Science Foundation of China (Grant No. LJXSZ26A040002). Computational resources were provided by the National Supercomputer Center in Tianjin. 

\bibliography{references}

\onecolumngrid
\clearpage
\begingroup
\setcounter{equation}{0}
\setcounter{table}{0}
\setcounter{figure}{0}
\renewcommand{\theequation}{S\arabic{equation}}
\renewcommand{\thetable}{S\Roman{table}}
\renewcommand{\thefigure}{S\arabic{figure}}
\begin{center}
\textbf{\large SUPPLEMENTAL MATERIAL FOR:}\\[0.5em]
\textbf{\large Quantum Metric Connects Bond Delocalization to Phonon Softening}\\[1em]
Tianqi Deng and Tiejun Zhu\\[0.5em]
\textit{State Key Laboratory of Silicon and Advanced Semiconductor Materials, School of Materials Science and Engineering, Zhejiang University, Hangzhou 310058, China}\\[0.5em]
\texttt{dengtq@zju.edu.cn}
\end{center}
\clearpage
\begin{bibunit}
\input{SM}
\begingroup
\section*{Supplementary References}
\makeatletter
\let\@FMN@list\@empty
\immediate\write\@bibunitaux{\string\citation{REVTEX42Control,apsrev42Control}}
\makeatother
\let\savedbiblabel\label
\renewcommand{\label}[1]{\savedbiblabel{SM-#1}}

\input{supplement-references.tex}
\endgroup
\end{bibunit}
\endgroup

\end{document}

%% file: SM.tex
\section{Computational procedures and data}

\begin{figure}[tbp]
 \includegraphics{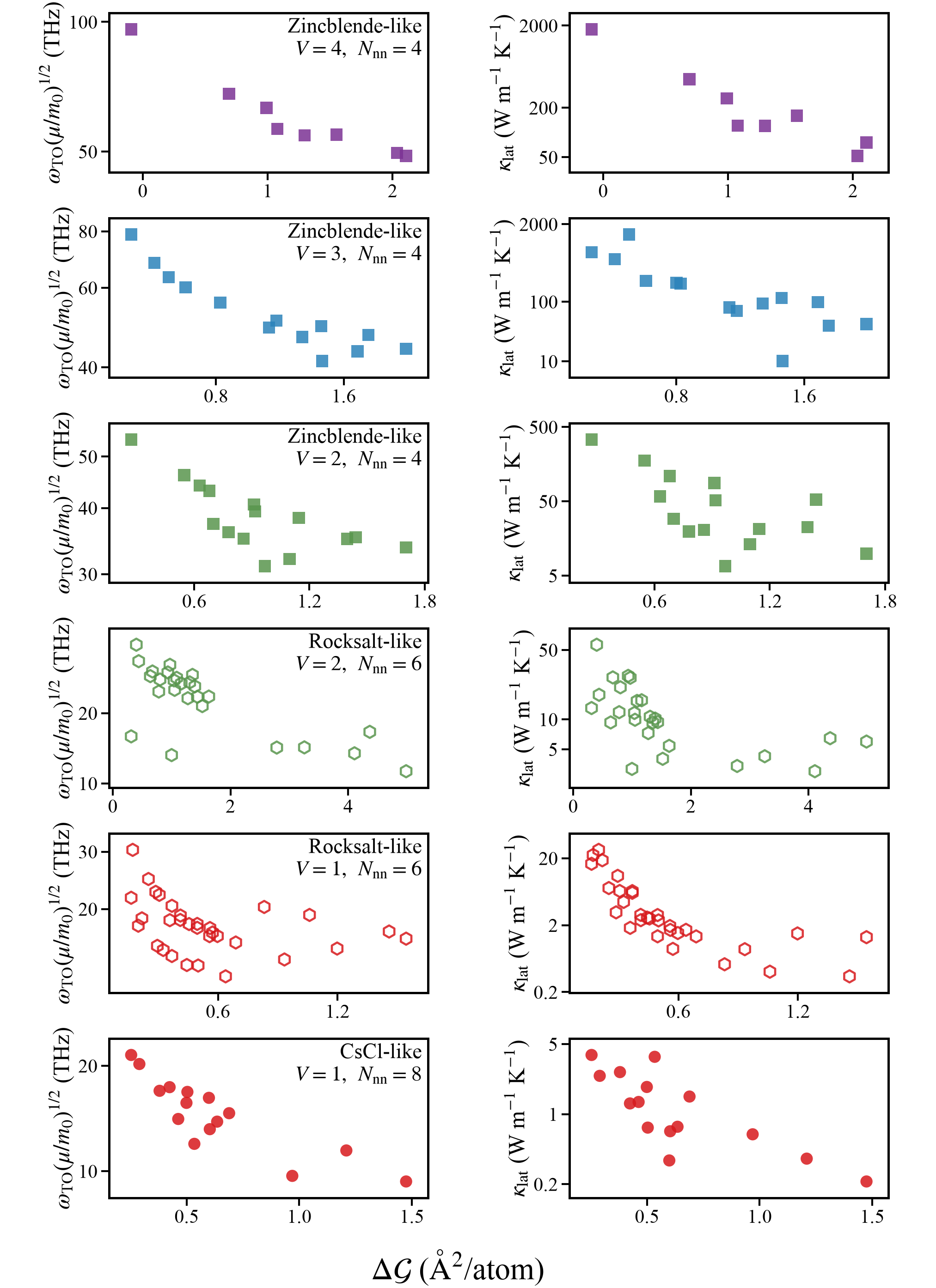}
 \caption{\label{fig_group}The mass-scaled transverse optical phonon frequency $\omega_{\mathrm{TO}} (\mu/m_0)^{1/2}$ (left) and lattice thermal conductivity $\kappa_{\mathrm{lat}}$ for representative valency and structure groups.}
\end{figure}

The frequency comparison in Fig.~\ref{fig_bond}(b) includes two-atom elemental and binary compounds in nominal-charge groups $V=$1--4, where elemental group-IV structures are also denoted by $V=$4. The detailed within-group comparison for representative valency, structure, and coordination number is shown in Fig.~\ref{fig_group}. The corresponding material data are listed in Table~\ref{tab:s1_materials}. Density-functional calculations use \texttt{VASP}~\cite{KresseFurthmuller1996CMS,KresseFurthmuller1996PRB} with projector augmented-wave datasets~\cite{KresseJoubert1999PAW} and the PBEsol functional~\cite{Perdew2008PBEsol}. We take the lowest zone-center optical frequency without the nonanalytic correction. This gives the TO frequency for cubic crystals. For the trigonal and hexagonal structures, the selected mode is transverse in the limit $\mathbf q\to 0$ along the $c$ axis. This frequency is positive for all structures in Fig.~\ref{fig_bond}(b). The conductivity comparison in Fig.~\ref{fig_bond}(c) also includes additional materials. Room-temperature lattice thermal conductivities for the database entries are taken from \texttt{Phonix}~\cite{Ohnishi2026Phonix}, generated by the automated \texttt{auto-kappa} workflow using \texttt{ALAMODE}~\cite{Tadano2014ALAMODE}. Additional room-temperature values for AgTlI$_2$~\cite{Zeng2024Thermal}, CsSnBr$_3$~\cite{Xie2020Halide}, and Bi$_2$O$_2$Se~\cite{Gibson2021Thermal} were obtained from measurements. The CsAg$_2$I$_3$ and CsCu$_2$I$_3$ values are estimated from Fig.~3(a) of Ref.~\cite{Shen2026Ultralow}. For Ge, we use the 300~K calculation for natural isotopic composition from Table~V of Ref.~\cite{McGaughey2025PhononOlympics}. For tetragonal BaTiO$_3$, we use the calculated in-plane conductivity at 300~K~\cite{Feger2024Thermal}. For PbTiO$_3$, we use the room-temperature single-crystal measurement~\cite{Tachibana2008Thermal}. The invariant spreads are obtained with \texttt{Wannier90}~\cite{Pizzi2020Wannier90} to evaluate the quantum metric. Nominal oxidation states are assigned with \texttt{pymatgen}~\cite{Ong2013Pymatgen}; the plotted valency is the magnitude of the nominal charge in these two-atom binary cells.

The Brillouin-zone measure in the main text is normalized per primitive cell, $[d\mathbf k]=V_{\mathrm{cell}}d^3k/(2\pi)^3$. Unless otherwise specified, $\mathcal G$ and $\Omega_I$ refer to one spatial occupied manifold, and spin multiplicity is included explicitly through $g_s$. For spin-degenerate calculations, $g_s=2$; when individual spinor states are summed, $g_s=1$. For the plotted descriptor, the summed spinor spread is divided by two to retain the same spatial-orbital convention. This normalization is applied before ionic subtraction and division by the atom count. We evaluate the isolated-ion spreads for nominal shell occupations using fixed PBEsol reference orbitals in a 20 Å cubic cell. Thus $\Delta\mathcal G$ is a scalar trace difference in units of length squared per atom.

The ferroelectric paths connect the nonpolar reference structures to the minimum-energy polar structures of GeTe, LiBeSb, and BaTiO$_3$. Their ionic references are Ge$^{2+}$/Te$^{2-}$, Li$^+$/Be$^{2+}$/Sb$^{3-}$, and Ba$^{2+}$/Ti$^{4+}$/O$^{2-}$, respectively. The Wannier cells contain two, six, and five atoms, respectively. GeTe and BaTiO$_3$ retain their reference lattice vectors, while the LiBeSb lattice parameters vary along the path. We use density-functional perturbation theory~\cite{Baroni2001,GonzeLee1997} in \texttt{VASP} to obtain the zone-center phonons and Born effective charges~\cite{Gajdos2006PAWResponse}. We track the polar mode of the symmetrized zone-center dynamical matrix using the mass-weighted distortion and eigenvector overlaps between adjacent structures, projecting onto the subspace for degenerate modes. Negative squared frequencies denote imaginary modes. The charge plots use the signed Cartesian $Z^*_{zz}$ for Te, Sb, and the O site on the Ti--O chain parallel to $z$ in the cubic reference, respectively. The pressure-dependent PbSe scan uses r2SCAN~\cite{Furness2020R2SCAN} with spin--orbit coupling~\cite{Steiner2016PAWSOC}. Conventional semilocal functionals can give incorrect band ordering in lead chalcogenides at ambient pressure~\cite{AguadoPuente2020Tellurides,Svane2010LeadChalcogenides}. The lines in Fig.~\ref{fig_topo}(b,c) are linear fits on either side of the crossover, excluding points near the gap minimum, and serve as guides to the eye.
\input{figure_S1_materials.tex}
\clearpage

\section{Born--Land\'e model for cubic ionic crystals}

We consider zincblende, rocksalt, and CsCl-type crystals in the rigid-ion framework of Born and Huang~\cite{BornHuang}. Each primitive cell contains two ions of charges $\pm Ze$ and masses $M_\pm$. Coulomb interactions act between all ions with a fixed permittivity $\epsilon$ ($\epsilon_0$ for unscreened ions). The repulsion $D/r^m$, with $D>0$ and $m>1$, acts only between nearest unlike ions separated by $r$.

Under uniform scaling, the energy per primitive cell and the zero-pressure equilibrium condition at $r=r_0$ are
\begin{align}
 E(r)&=-\frac{\alpha C}{r}+\frac{N_{\mathrm{nn}}D}{r^m},\qquad C=\frac{Z^2e^2}{4\pi\epsilon},\label{eq:cell-energy}\\
 E'(r_0)&=0,\qquad N_{\mathrm{nn}}D=\frac{\alpha C}{m}r_0^{m-1}.\label{eq:equilibrium}
\end{align}
Here $N_{\mathrm{nn}}$ is the number of nearest unlike-ion neighbors per ion, and $\alpha$ is the Madelung constant evaluated by Ewald summation~\cite{Ewald1921LatticeSums}. The equilibrium primitive-cell volume is $V_{\mathrm{cell}}=cr_0^3$ (Table~\ref{tab:structures}). The bulk modulus is $B_0=\alpha(m-1)C/(9cr_0^4)$, and the bulk elastic speed scale is $v_B=\sqrt{B_0/\rho}$, with $\rho=(M_++M_-)/V_{\mathrm{cell}}$.

The optical modes involve the relative displacement $\mathbf w=\mathbf u_+-\mathbf u_-$ with reduced mass $\mu=M_+M_-/(M_++M_-)$. For a central potential $V(r)$, the bond Hessian is
\begin{equation}
 \boldsymbol\Phi^{(V)}(\mathbf r)=\left(V''-\frac{V'}{r}\right)\widehat{\mathbf r}\widehat{\mathbf r}^{\mathsf T}+\frac{V'}{r}\mathbf I,
 \qquad \widehat{\mathbf r}=\frac{\mathbf r}{r}.
 \label{eq:bond-ifc}
\end{equation}
Cubic symmetry gives $\sum_j\widehat{\mathbf r}_j\widehat{\mathbf r}_j^{\mathsf T}=(N_{\mathrm{nn}}/3)\mathbf I$ for the nearest-neighbor shell. Summing the Hessians of $V_{\mathrm{SR}}=D/r^m$ and using Eq.~\eqref{eq:equilibrium} gives the isotropic stiffness
\begin{equation}
 \Phi_{\mathrm{SR}}=\frac{N_{\mathrm{nn}}}{3}\left[V_{\mathrm{SR}}''(r_0)+\frac{2V_{\mathrm{SR}}'(r_0)}{r_0}\right]
 =\frac{\alpha(m-1)C}{3r_0^3}.
 \label{eq:repulsive-stiffness}
\end{equation}
The transverse term remains because $V_{\mathrm{SR}}'(r_0)$ is finite at crystal equilibrium.

For the Coulomb contribution, consider a spherical Lorentz cavity much larger than the lattice spacing and much smaller than the phonon wavelength. Each complete shell of either sublattice gives zero linear field because the Hessian of $1/r$ has zero trace and its cubic shell sum is isotropic. The cavity surface charge $-\mathbf P\cdot\widehat{\mathbf r}$ gives a field $\mathbf P/(3\epsilon)$ at its center, using the spherical average $\langle\widehat{\mathbf r}\widehat{\mathbf r}^{\mathsf T}\rangle=\mathbf I/3$. With the polarization $\mathbf P=Ze\mathbf w/V_{\mathrm{cell}}$, the local field including the macroscopic longitudinal field from Gauss's law is
\begin{equation}
 \boldsymbol{\mathcal E}_{\mathrm{loc}}=\frac{\mathbf P}{3\epsilon}-\frac{\widehat{\mathbf q}(\widehat{\mathbf q}\cdot\mathbf P)}{\epsilon}.
 \label{eq:lorentz}
\end{equation}
The relative motion obeys $\mu\ddot{\mathbf w}=-\Phi_{\mathrm{SR}}\mathbf w+Ze\boldsymbol{\mathcal E}_{\mathrm{loc}}$. For $\mathbf w\perp\widehat{\mathbf q}$, the macroscopic field vanishes and the two degenerate TO modes have
\begin{equation}
 \omega_{\mathrm{TO}}^2=\frac{1}{\mu}\left(\Phi_{\mathrm{SR}}-\frac{Z^2e^2}{3\epsilon V_{\mathrm{cell}}}\right)
 =\frac{C}{3\mu r_0^3}\left[\alpha(m-1)-\frac{4\pi}{c}\right].
 \label{eq:TO}
\end{equation}
The longitudinal optical (LO) mode has $\omega_{\mathrm{LO}}^2=\omega_{\mathrm{TO}}^2+Z^2e^2/(\mu\epsilon V_{\mathrm{cell}})$. Optical stability requires $\alpha c(m-1)>4\pi$.

At fixed nearest-neighbor distance, increasing coordination from zincblende through rocksalt to CsCl type reduces the primitive-cell volume. For the same ionic charge and relative displacement, the smaller volume gives a larger polarization and a stronger Lorentz-field reduction of the transverse stiffness.

\begin{table}[t]
\caption{Geometry and Madelung constants for the three cubic structures. Here $N_{\mathrm{nn}}$ is the coordination number, $a$ is the conventional cubic lattice parameter, and $V_{\mathrm{cell}}=cr_0^3$.}
\label{tab:structures}
\begin{ruledtabular}
\begin{tabular}{@{}lcccc@{}}
Structure & $N_{\mathrm{nn}}$ & $r_0/a$ & $c$ & $\alpha$ \\
Zincblende & 4 & $\sqrt{3}/4$ & $16/(3\sqrt{3})$ & 1.63806 \\
Rocksalt & 6 & $1/2$ & 2 & 1.74756 \\
CsCl type & 8 & $\sqrt{3}/2$ & $8/(3\sqrt{3})$ & 1.76267 \\
\end{tabular}
\end{ruledtabular}
\end{table}

\section{Force-constant changes from charge redistribution}
\label{sec:sm:charge-redistribution}

Conventional density-functional perturbation theory treats the nuclei, or the nuclei together with frozen core electrons, separately from the responding electrons. In the rigid-ion reference used here, each ion additionally carries a reference valence-electron density that moves rigidly with it. We take this rigid-ion crystal as a reference and calculate the IFC change due to charge redistribution at the same ionic geometry.

Let $U^{\mathrm{ion}}$ be the rigid-ion energy, including short-range repulsion. At fixed strain and electrical boundary conditions, the model energy for a trial density change $\Delta\tilde n$ is
\begin{equation}
 U[\Delta\tilde n,\{\boldsymbol\tau\}]
 =U^{\mathrm{ion}}(\{\boldsymbol\tau\})
 +\int \Delta\tilde n(\mathbf r)v^{\mathrm{ion}}(\mathbf r)\,d\mathbf r
 +E_{\mathrm{el}}[\Delta\tilde n].
 \label{eq:sm:redistribution-energy}
\end{equation}
Here $v^{\mathrm{ion}}=\sum_\kappa v_{\kappa}^{\mathrm{ion}}(\mathbf r-\boldsymbol\tau_\kappa)$ is the ionic potential energy coupled to the density redistribution in this model. It represents the interaction of the redistribution with the rigid-ion reference. Using this model potential in the microscopic response requires a consistent choice of reference and coupling; the net-charge estimate below is an additional approximation. The translationally invariant functional $E_{\mathrm{el}}$ stabilizes the density minimum and has no explicit dependence on ion positions at fixed $\Delta\tilde n$. For a reference with the crystal electron count, $\int\Delta\tilde n\,d\mathbf r=0$. Denote the minimizing density change by $\Delta n=n-\sum_\kappa n_{\kappa}^{\mathrm{ion}}$, where $n$ is the ground-state density.

At the minimum, $\delta U/\delta(\Delta\tilde n)=\lambda$ is spatially constant. The implicit density derivative contributes $\lambda\int\partial_{\kappa\alpha}\Delta n\,d\mathbf r=0$ to the first energy derivative. With $\partial_{\kappa\alpha}=\partial/\partial\tau_{\kappa\alpha}$, we obtain
\begin{align}
 F_{\kappa\alpha}
 &=-\partial_{\kappa\alpha}U^{\mathrm{ion}}
 -\int \Delta n\,\partial_{\kappa\alpha}v_{\kappa}^{\mathrm{ion}}\,d\mathbf r,
 \label{eq:sm:redistribution-force}\\
 \Phi_{\kappa\alpha,\kappa'\beta}-\Phi^{\mathrm{ion}}_{\kappa\alpha,\kappa'\beta}
 &=\int (\partial_{\kappa\alpha}\Delta n)
          (\partial_{\kappa'\beta}v_{\kappa'}^{\mathrm{ion}})\,d\mathbf r\nonumber\\
 &\quad+\delta_{\kappa\kappa'}\int \Delta n\,
          \partial_{\kappa\alpha}\partial_{\kappa\beta}v_{\kappa}^{\mathrm{ion}}\,d\mathbf r,
 \label{eq:sm:redistribution-ifc}
\end{align}
Here $\Phi^{\mathrm{ion}}$ is the Hessian of $U^{\mathrm{ion}}$. For distinct ions, only the first term contributes. Including the Coulomb self-energy in $E_{\mathrm{el}}$ generally changes $\Delta n$ and its response while preserving the form of Eq.~\eqref{eq:sm:redistribution-ifc}.

The full crystal and the rigid-ion reference are each invariant under a uniform translation. The acoustic sum rule therefore fixes their difference on site for each pair of Cartesian directions:
\begin{equation}
 \Phi_{\kappa\alpha,\kappa\beta}-\Phi^{\mathrm{ion}}_{\kappa\alpha,\kappa\beta}
 =-\sum_{\kappa'\ne\kappa}
 \left(\Phi_{\kappa\alpha,\kappa'\beta}-\Phi^{\mathrm{ion}}_{\kappa\alpha,\kappa'\beta}\right).
 \label{eq:sm:redistribution-asr}
\end{equation}
The sum includes all other ions, including periodic images. The cross-ion IFC changes thus determine the on-site change.

To connect the density response to Eq.~\eqref{eq:main:ifc-response}, write the total electron density as the sum of the occupied Bloch-state probability densities. For the zone-center displacement used in that equation, all translational copies of atom $\kappa$ move together, while the occupations and the cell volume remain fixed. Differentiating the sum gives
\begin{align}
 n(\mathbf r)
 &=g_s\sum_{n\in\mathrm{occ}}\int[d\mathbf k]\,
 u_{n\mathbf k}^{\dagger}(\mathbf r)u_{n\mathbf k}(\mathbf r),
 \label{eq:sm:occupied-density}\\
 \partial_{\kappa\alpha}n(\mathbf r)
 &=2g_s\sum_{n\in\mathrm{occ}}\int[d\mathbf k]\,
 \operatorname{Re}\!\left[
 u_{n\mathbf k}^{(\tau_{\kappa\alpha})\dagger}(\mathbf r)
 u_{n\mathbf k}(\mathbf r)\right].
 \label{eq:sm:density-derivative}
\end{align}
The factor of two comes from differentiating the orbital and its complex conjugate. Rotations among occupied states leave the summed density unchanged, so the derivative can be evaluated using the displacement tangent projected outside the occupied manifold. Multiplying by the ionic perturbation and integrating over one primitive cell gives the zone-center response
\begin{equation}
\begin{aligned}
 &\int\partial_{\kappa\alpha}n \partial_{\kappa'\beta}v^{\mathrm{ion}} \,d\mathbf r\\
 &\qquad=
 2g_s\sum_{n\in\mathrm{occ}}\int[d\mathbf k]\,
 \operatorname{Re}\langle u_{n\mathbf k}^{(\tau_{\kappa\alpha})}|
 \partial_{\kappa'\beta}v^{\mathrm{ion}}|u_{n\mathbf k}\rangle
 =\Phi^{\mathrm{resp}}_{\kappa\alpha,\kappa'\beta}.
\end{aligned}
 \label{eq:sm:density-response-ifc}
\end{equation}
This is Eq.~\eqref{eq:main:ifc-response}, whose geometric approximation is given by Eq.~\eqref{eq:main:ifc-qgt}. The zone-center response includes contributions from ion pairs in different cells. For distinct sublattices, the correction $\Delta\Phi=\Phi-\Phi^{\mathrm{ion}}$ follows by subtracting the response of the rigid-ion reference:
\begin{equation}
 \Phi_{\kappa\alpha,\kappa'\beta}-\Phi^{\mathrm{ion}}_{\kappa\alpha,\kappa'\beta}
 =\Phi^{\mathrm{resp}}_{\kappa\alpha,\kappa'\beta}
 -\int \partial_{\kappa\alpha}n_{\kappa}^{\mathrm{ion}}
       \partial_{\kappa'\beta}v^{\mathrm{ion}}\,d\mathbf r
 \qquad(\kappa\ne\kappa').
 \label{eq:sm:redistribution-subtraction}
\end{equation}
Here $n_{\kappa}^{\mathrm{ion}}$ is the periodic sum of the reference densities of ion $\kappa$; the integral over one primitive cell is the response from their rigid translation. At the geometric level, representing the ionic reference by additive isolated-ion metrics and using the same effective coefficient $C_{\kappa\alpha,\kappa'\beta}$ for both responses gives the metric difference $\mathcal G^{\alpha\beta}-\sum_\kappa \mathcal G_{\kappa}^{\mathrm{ion},\alpha\beta}$. Its spin-normalized trace per atom is the descriptor $\Delta\mathcal G$. The IFC subtraction generally retains the force-gradient and excitation-energy weights of the crystal and reference separately.

\section{Wave-function tangents and the geometric contribution}

The crystal response $\Phi^{\mathrm{resp}}$ in Eq.~\eqref{eq:main:ifc-response} pairs the density response to one ionic displacement with the ionic potential derivative for the other. The density response is generated by the self-consistent perturbation, which includes electronic screening, while the force is evaluated using the ionic potential derivative. We first obtain the wave-function response to the screened perturbation, then expand both displacement potentials locally to first order to connect $\Phi^{\mathrm{resp}}$ to the quantum metric.

Define the occupied projector $P_{\mathbf k}=\sum_{n\in\mathrm{occ}}|u_{n\mathbf k}\rangle\langle u_{n\mathbf k}|$. In the parallel-transport gauge, the displacement tangent satisfies
\begin{equation}
 (1-P_{\mathbf k})(H_{\mathbf k}-E_{n\mathbf k})(1-P_{\mathbf k})|u_{n\mathbf k}^{(\tau_{\kappa\alpha})}\rangle
 =-(1-P_{\mathbf k})\partial_{\kappa\alpha}v_{\mathrm{scf}}|u_{n\mathbf k}\rangle,
 \label{eq:sm:sternheimerph}
\end{equation}
where $v_{\mathrm{scf}}$ is the screened self-consistent field (SCF) potential. This gives
\begin{equation}
 \langle u_{m\mathbf k}|u_{n\mathbf k}^{(\tau_{\kappa\alpha})}\rangle
 =-\frac{g_{mn,\mathbf k}^{\kappa\alpha}}{\Delta_{mn\mathbf k}},\qquad
 g_{mn,\mathbf k}^{\kappa\alpha}=\langle u_{m\mathbf k}|\partial_{\kappa\alpha}v_{\mathrm{scf}}|u_{n\mathbf k}\rangle.
\end{equation}
Here $m$ is unoccupied and $\Delta_{mn\mathbf k}=E_{m\mathbf k}-E_{n\mathbf k}>0$.

Following the leading-order displacement-potential expansion of Ref.~\cite{Hu2025EPCGeometry}, we expand $\partial v$ to first order for $v=v^{\mathrm{ion}}$ or $v_{\mathrm{scf}}$ about $\mathbf r_0$ where the states overlap most strongly
\begin{equation}
\partial v(\mathbf r)\simeq \partial v(\mathbf r_0)
+(\mathbf r-\mathbf r_0)\cdot
\left.\nabla_{\mathbf r}\partial v\right|_{\mathbf r_0}.
\end{equation}
The constant term vanishes by orthogonality. The linear term couples $\nabla_{\mathbf r}\partial v$ at $\mathbf r_0$ to the interband position matrix element $i\langle u_{m\mathbf k}|\nabla_{\mathbf k}u_{n\mathbf k}\rangle$. When the component along the ionic displacement dominates this contraction, the ionic and screened matrix elements take the following forms, with $\gamma$ and $\tilde\gamma$ the corresponding gradient components $\partial_{r_\alpha} \partial_{\kappa\alpha} v$
\begin{align}
 \langle u_{m\mathbf k}|\partial_{\kappa\alpha}v^{\mathrm{ion}}|u_{n\mathbf k}\rangle
 &\simeq\gamma_{mn,\mathbf k}^{\kappa\alpha}\langle u_{m\mathbf k}|i\partial_{k_\alpha}u_{n\mathbf k}\rangle,\label{eq:sm:bare-coupling}\\
 g_{mn,\mathbf k}^{\kappa\alpha}
 &\simeq\tilde\gamma_{mn,\mathbf k}^{\kappa\alpha}\langle u_{m\mathbf k}|i\partial_{k_\alpha}u_{n\mathbf k}\rangle.
 \label{eq:sm:screened-coupling}
\end{align}

Inserting the tangent and both couplings into the IFC gives
\begin{align}
 \Phi^{\mathrm{resp}}_{\kappa\alpha,\kappa'\beta}
 &\simeq-2g_s\sum_{\substack{n\in\mathrm{occ}\\m\notin\mathrm{occ}}}\int[d\mathbf k]\,
 \frac{\tilde\gamma_{mn,\mathbf k}^{\kappa\alpha}\gamma_{mn,\mathbf k}^{\kappa'\beta}}{\Delta_{mn\mathbf k}}
 \operatorname{Re}Q_{mn\mathbf k}^{\alpha\beta}\nonumber\\
 &\simeq-\frac{2g_s\tilde\gamma_{\mathrm{eff}}^{\kappa\alpha}\gamma_{\mathrm{eff}}^{\kappa'\beta}}{\Delta_{\mathrm{eff}}}\mathcal G^{\alpha\beta}.
 \label{eq:sm:ifc-qgt}
\end{align}
The first line retains the transition-dependent weight $\tilde\gamma\gamma/\Delta$. In the second line, the effective coefficient represents this weight for the specified metric component.

\section{Sign of the response}

To first order, the IFC difference $\Delta\Phi=\Phi-\Phi^{\mathrm{ion}}$ shifts the squared frequency of a rigid-ion mode with normalized mass-weighted eigenvector $\mathbf e_\nu$ by
\begin{equation}
 \Delta\omega_\nu^2\simeq
 \mathbf e_\nu^\dagger M^{-1/2}\Delta\Phi M^{-1/2}\mathbf e_\nu,
 \label{eq:sm:mode}
\end{equation}
where $M$ is the diagonal atomic-mass matrix. The alternating-chain illustration below determines the sign of the optical-mode projection.

To illustrate the sign of the force gradient near a bond, we approximate the effective potential of each rigid reference ion by the electrostatic potential of a point charge $Z_\kappa^{\mathrm{ion}}e$:
\begin{equation}
 v_\kappa^{\mathrm{ion}}(r)\simeq-\frac{Z_\kappa^{\mathrm{ion}}e^2}{4\pi\epsilon_0r}.
\end{equation}
This effective-ion approximation includes the reference ion's net charge and is used only for the sign estimate. Its displacement derivative and local gradient are
\begin{align}
 \partial_{\tau_{\kappa\alpha}}v_\kappa^{\mathrm{ion}}
 &\simeq-\frac{e^2Z_\kappa^{\mathrm{ion}}}{4\pi\epsilon_0}\frac{r_\alpha}{r^3},\\
 \partial_{r_\beta}\partial_{\tau_{\kappa\alpha}}v_\kappa^{\mathrm{ion}}
 &\simeq\frac{e^2Z_\kappa^{\mathrm{ion}}}{4\pi\epsilon_0}
 \frac{3r_\alpha r_\beta-\delta_{\alpha\beta}r^2}{r^5}.
\end{align}
Consider an alternating chain with $Z_{\mathrm{cation}}^{\mathrm{ion}}>0$ and $Z_{\mathrm{anion}}^{\mathrm{ion}}<0$. Along the bond between neighboring ions, the point-charge estimate gives longitudinal force gradients of opposite sign:
\begin{align}
\gamma_{\mathrm{+}} &\simeq \frac{e^2Z_{\mathrm{cation}}^{\mathrm{ion}}}{4\pi\epsilon_0}\frac{2}{r^3} > 0,\\
\gamma_{\mathrm{-}} &\simeq \frac{e^2Z_{\mathrm{anion}}^{\mathrm{ion}}}{4\pi\epsilon_0}\frac{2}{r^3} < 0.
\end{align}
If screening preserves the gradient signs, the nearest-neighbor coefficient satisfies $C_{\mathrm{nn}}<0$. Using the same coefficient for the crystal and rigid-ion reference, the longitudinal metric gain $\Delta\mathcal G^{xx}>0$ gives a positive IFC correction:
\begin{equation}
\Delta\Phi_{\mathrm{nn}} \simeq -\frac{2g_s\tilde\gamma_{\mathrm{eff}}^{\mathrm{+}}\gamma_{\mathrm{eff}}^{\mathrm{-}}}{\Delta_{\mathrm{eff}}}\Delta\mathcal G^{xx} > 0.
\end{equation}
The two equivalent nearest-neighbor bonds give an off-diagonal correction of $2\Delta\Phi_{\mathrm{nn}}$ at $\Gamma$, so the acoustic sum rule fixes each on-site correction to $-2\Delta\Phi_{\mathrm{nn}}$. The displacement pattern $(1/2,-1/2)$ gives a unit relative displacement; a uniform translation does not change the projected stiffness by the acoustic sum rule. The resulting change in squared optical frequency relative to the rigid-ion reference is
\begin{equation}
\Delta\omega^2_{\mathrm{O}} = \frac{1}{\mu} \begin{pmatrix} \frac{1}{2} & -\frac{1}{2} \end{pmatrix}
\begin{pmatrix}
-2\Delta\Phi_{\mathrm{nn}} & 2\Delta\Phi_{\mathrm{nn}} \\
2\Delta\Phi_{\mathrm{nn}} & -2\Delta\Phi_{\mathrm{nn}}
\end{pmatrix}
\begin{pmatrix}
\frac{1}{2} \\ -\frac{1}{2}
\end{pmatrix} = -\frac{2}{\mu} \Delta\Phi_{\mathrm{nn}} < 0.
\end{equation}
The negative shift in the squared frequency gives the optical softening in Eq.~\eqref{eq:main:mode}.

\section{Quantum geometry of the one-dimensional SSH model}

Consider a half-filled SSH chain~\cite{Su1979SSH} with orthonormal orbitals and equal on-site energies set to zero. The short and long bonds have lengths $l-\Delta$ and $l+\Delta$ and hoppings $t+\eta$ and $t-\eta$, with $0<\Delta<l$, $0<\eta<t$, and the short bond inside the cell. The cell length is $2l$, and the normalized per-cell measure is $[dk]=(l/\pi)\,dk$ in the Brillouin zone $\mathrm{BZ}=[-\pi/(2l),\pi/(2l))$.

In the periodic unit-cell Fourier convention $\hat c_{j\alpha}=N^{-1/2}\sum_k e^{2ikjl}\hat c_{k\alpha}$, where $\alpha=A,B$ and $N$ is the number of cells, the Bloch Hamiltonian and its eigenvalues are
\begin{align}
 H(k)&=\begin{pmatrix}0&f(k)\\ f(k)^*&0\end{pmatrix},
 \qquad f(k)=(t+\eta)+(t-\eta)e^{-2ikl},
 \label{eq:ssh_H}\\
 E_\pm(k)&=\pm2\sqrt{D(k)},
 \qquad D(k)=t^2\cos^2(kl)+\eta^2\sin^2(kl).
 \label{eq:ssh_energy}
\end{align}
In this convention, the positions of the orbitals within the cell are specified separately from the hopping matrix~\cite{FuchsPiechon2021Embedding}.

Writing $f(k)=|f(k)|e^{i\phi(k)}$, we choose spinors $|u_n(k)\rangle=(n e^{i\phi(k)},1)^{\mathsf T}/\sqrt{2}$ for $n=\pm$. Using $f(k)=2e^{-ikl}[t\cos(kl)+i\eta\sin(kl)]$, the phase derivative and the periodic-basis spinor metric are
\begin{align}
 \phi'(k)&=\frac{\operatorname{Im}[f(k)^*\partial_k f(k)]}{|f(k)|^2}
 =l\left[\frac{t\eta}{D(k)}-1\right],\\
 \mathcal G_k&=\langle\partial_k u_n|\partial_k u_n\rangle
 -|\langle u_n|\partial_k u_n\rangle|^2
 =\frac{[\phi'(k)]^2}{4}.
 \label{eq:ssh_metric_def}
\end{align}
The substitution $u=\tan(kl)$ gives the first integral below. Differentiating it with respect to $t$ and $\eta$ gives the second,
\begin{equation}
 \int\frac{[dk]}{D(k)}=\frac{1}{t\eta},
 \qquad
 \int\frac{[dk]}{D(k)^2}=\frac{t^2+\eta^2}{2t^3\eta^3}.
\end{equation}
The occupied-band metric per cell is therefore
\begin{equation}
 \mathcal G_{\mathrm{SSH}}=\int[dk]\,\frac{[\phi'(k)]^2}{4}
 =\frac{l^2(t-\eta)^2}{8t\eta}.
 \label{eq:ssh_integral}
\end{equation}
At fixed $t$ and $l$, the gap $4\eta$ closes and $\mathcal G_{\mathrm{SSH}}$ diverges as $\eta\to0^+$. Including the intracell orbital positions in the point-orbital model adds a bounded term to the phase derivative, leaving this divergence unchanged. If both hoppings follow the same strictly monotonic function of bond length, this is the equal-bond limit.

\section{Dielectric response and Born effective charges from charge redistribution}
\label{sec:sm:dielectric-bec-redistribution}

We express all charges denoted by $Z$ in units of $e$. The Born effective charge (BEC) is then $Z^*_{\kappa\alpha,\beta}=(V_{\mathrm{cell}}/e)(\partial P_\beta/\partial\tau_{\kappa\alpha})_{\boldsymbol{\mathcal E}=0}$~\cite{GonzeLee1997}. The anomaly $\Delta Z_{\kappa\alpha,\beta}=Z^*_{\kappa\alpha,\beta}-Z_\kappa^{\mathrm{ion}}\delta_{\alpha\beta}$ is given by the dipole of the density difference $\Delta n$ induced by displacing ion $\kappa$ in the reference cell:
\begin{equation}
 \Delta Z_{\kappa\alpha,\beta}
 =-\int_{\mathbb R^3} r_\beta
 \frac{\partial\Delta n(\mathbf r)}{\partial\tau_{0\kappa\alpha}}\,d^3r .
 \label{eq:sm:bec-redistribution-moment}
\end{equation}
The response is evaluated at zero macroscopic electric field. With macroscopic screening included, the first moment of the total charge response gives the Callen charge for an isotropic dielectric~\cite{HongVanderbilt2013Flexoelectricity}.

For the crystal density $n$, summing the Wannier-orbital responses to the single-ion displacement is equivalent by translation symmetry to taking the zone-center displacement response of the Wannier orbitals in the reference cell. The interband position matrix element in the Bloch representation is $i\langle u_{m\mathbf k}|\partial_{k_\beta}u_{n\mathbf k}\rangle$. Let $Z_\kappa^0$ be the charge of the nucleus plus frozen core, in units of $e$. The valence-electron contribution $Z^{\mathrm{val}}_{\kappa\alpha,\beta}=Z^*_{\kappa\alpha,\beta}-Z_\kappa^0\delta_{\alpha\beta}$ takes the nonstationary form of Gonze and Lee [Eq.~(42) of Ref.~\cite{GonzeLee1997}]. Inserting the screened displacement coupling of Eq.~\eqref{eq:sm:screened-coupling} gives
\begin{align}
 Z^{\mathrm{val}}_{\kappa\alpha,\beta}
 &=-2g_s\sum_{n\in\mathrm{occ}}\int[d\mathbf k]\,
 \operatorname{Re}\langle u_{n\mathbf k}^{(\tau_{\kappa\alpha})}|i\partial_{k_\beta}u_{n\mathbf k}\rangle
 \label{eq:sm:bec-redistribution-bloch}\\
 &\simeq 2g_s\sum_{\substack{n\in\mathrm{occ}\\m\notin\mathrm{occ}}}\int[d\mathbf k]\,
 \frac{\tilde\gamma_{mn,\mathbf k}^{\kappa\alpha}\operatorname{Re}Q_{mn\mathbf k}^{\alpha\beta}}{\Delta_{mn\mathbf k}}
 \simeq \frac{2g_s\tilde\gamma_{\mathrm{eff}}^{\kappa\alpha}}{\Delta_{\mathrm{eff}}}\mathcal G^{\alpha\beta}.
 \label{eq:sm:bec-redistribution-metric}
\end{align}
The reference valence density carries charge $Z_\kappa^{\mathrm{ion}}-Z_\kappa^0$. Subtracting its rigid-translation contribution gives the anomaly,
\begin{equation}
 \Delta Z_{\kappa\alpha,\beta}
 =Z^{\mathrm{val}}_{\kappa\alpha,\beta}
 -(Z_\kappa^{\mathrm{ion}}-Z_\kappa^0)\delta_{\alpha\beta}.
\end{equation}

The relative high-frequency dielectric tensor $\epsilon^\infty$ follows from the electric-field tangent. Neglecting microscopic local-field effects, the field couples through $e r_\alpha$, and the tangent obeys
\begin{equation}
 (1-P_{\mathbf k})(H_{\mathbf k}-E_{n\mathbf k})(1-P_{\mathbf k})|u_{n\mathbf k}^{(\mathcal E_\alpha)}\rangle
 =-e(1-P_{\mathbf k})|i\partial_{k_\alpha}u_{n\mathbf k}\rangle.
 \label{eq:sm:dielectric-redistribution-sternheimer}
\end{equation}
Projecting onto an unoccupied state $u_{m\mathbf k}$ gives $-e\langle u_{m\mathbf k}|i\partial_{k_\alpha}u_{n\mathbf k}\rangle/\Delta_{mn\mathbf k}$. Substituting this component into the field-tangent expression for the dielectric tensor gives
\begin{align}
 \epsilon^\infty_{\alpha\beta}-\delta_{\alpha\beta}
 &\simeq -\frac{2g_s e}{\epsilon_0 V_{\mathrm{cell}}}\sum_{n\in\mathrm{occ}}\int[d\mathbf k]\,
 \operatorname{Re}\langle u_{n\mathbf k}^{(\mathcal E_\alpha)}|i\partial_{k_\beta}u_{n\mathbf k}\rangle
 \label{eq:sm:dielectric-redistribution-tangent}\\
 &\simeq \frac{2g_s e^2}{\epsilon_0 V_{\mathrm{cell}}}
 \sum_{\substack{n\in\mathrm{occ}\\m\notin\mathrm{occ}}}\int[d\mathbf k]\,
 \frac{\operatorname{Re}Q_{mn\mathbf k}^{\alpha\beta}}{\Delta_{mn\mathbf k}}
 \simeq \frac{2g_s e^2}{\epsilon_0 V_{\mathrm{cell}}\Delta_{\mathrm{eff}}}\mathcal G^{\alpha\beta}.
 \label{eq:sm:dielectric-redistribution-metric}
\end{align}
Here $\epsilon_0$ is the vacuum permittivity. At fixed strain, ionic relaxation adds a lattice contribution to the relative static dielectric tensor $\epsilon^s$~\cite{WuVanderbiltHamann2005},
\begin{equation}
 \epsilon^s_{\alpha\beta}-\epsilon^\infty_{\alpha\beta}
 = \frac{e^2}{\epsilon_0 V_{\mathrm{cell}}}\sum_{\kappa a,\kappa'b}
 Z^*_{\kappa a,\alpha}(\Phi^+)_{\kappa a,\kappa'b}Z^*_{\kappa'b,\beta},
 \label{eq:sm:dielectric-redistribution-static}
\end{equation}
where $\Phi^+$ is the inverse of the zone-center IFC matrix on the stable optical subspace and vanishes on rigid translations.

\section{Wannier bond orbitals and the RD--RO map}

WBOs span the occupied manifold. Their centers resolve local electronic polarization and have established connections to bulk polarization, oxidation states, and topology~\cite{KingSmith1993Polarization,Jiang2012OxidationStates,Yu2011TopologicalInvariant}. The Wannier gauge affects the final WBO ordering and shapes but preserves the quantum metric and invariant spread sum. For all materials investigated in this work except for the ferroelectric examples (GeTe, LiBeSb, BaTiO$_3$), we use unconstrained maximally localized Wannier functions~\cite{Marzari2012MLWF}, initialized with selected columns of the density matrix (SCDM)~\cite{Damle2018Disentanglement,Damle2017SCDMk}. For the ferroelectric examples in Fig.~\ref{fig_fe}, to preserve the WBO order and consistently track its change, we use atomic initial guesses for Wannierization and track the WBO corresponding to the same anion's $p_z$ orbitals.

For a normalized orbital $w_m$, define its covariance tensor
\begin{equation}
 \Omega_{\alpha\beta}^{(m)}=\langle w_m|r_\alpha r_\beta|w_m\rangle
 -\langle w_m|r_\alpha|w_m\rangle\langle w_m|r_\beta|w_m\rangle .
 \label{eq:sm:covariance}
\end{equation}
Its largest eigenvalue $\lambda_{\max}$ sets the principal extent,
\begin{equation}
 L_{\mathrm{WBO}}=\sqrt{3\lambda_{\max}},\qquad
 L_{\mathrm{Bader}}=\frac{V_{\mathrm{Bader}}^{1/3}}{2},\qquad
 \mathrm{RD}=\frac{ L_{\mathrm{WBO}}}{L_{\mathrm{Bader}}}.
 \label{eq:sm:rd}
\end{equation}
The volume is obtained by Bader partitioning~\cite{Bader1990Atoms} with the \texttt{bader} program~\cite{Henkelman2006Bader,Tang2009Bader} and belongs to the atom nearest the WBO center. The factor $\sqrt{3}$ makes $L_{\mathrm{WBO}}$ equal to the square root of the total orbital spread in the isotropic limit, while retaining sensitivity to the longest principal axis for directional orbitals. The Bader length is the volume-based convention used in the analysis, not an equal-volume spherical radius.

Let $L_{\mathrm{1W}}$ be the distance between the WBO center and the nearest atom, and $L_{\mathrm{12}}$ the separation of the nearest and second-nearest atoms, including periodic images. Relative offcenterness is
\begin{equation}
 \mathrm{RO}=\frac{L_{\mathrm{1W}}}{L_{\mathrm{12}}}.
 \label{eq:sm:ro}
\end{equation}
An atom-centered orbital has RO $=0$, while a midpoint-centered bond orbital has RO $=1/2$. Each material in Fig.~\ref{fig_rdro}(b) is represented by the WBO with the largest RD.

The invariant spread $\Omega_I$ is independent of rotations within the occupied manifold. Individual WBO spreads and centers resolve the chemistry in a specified localized gauge, rather than uniquely partitioning $\Omega_I$ into independent bonds. Symmetry-adapted localization~\cite{Sakuma2013SymmetryAdapted} can pin orbital centers to atomic or bond sites, whereas unconstrained localization can shift them away from high-symmetry positions while reducing the total spread. We therefore use the same localization convention for comparisons in the main text.

\clearpage

%% file: figure_S1_materials.tex
\begin{table}[p]
\centering
\caption{Materials selected for Fig.~\ref{fig_bond}(b) and Fig.~\ref{fig_group}. $\Delta\mathcal{G}$ is in \AA$^2$/atom, frequencies are in THz, and $\kappa_{\rm lat}$ is in W\,m$^{-1}$\,K$^{-1}$. Wurtzite GaN is included in the conductivity panel of Fig.~\ref{fig_group}.}
\label{tab:s1_materials}
\small
\setlength{\tabcolsep}{5pt}
\begin{tabular*}{\textwidth}{@{\extracolsep{\fill}}llrrrrrr}
\toprule
Formula & Space group & $V$ & $N_{\rm nn}$ & $\Delta\mathcal{G}$ & $\omega_{\rm TO}$ & $\omega_{\rm TO}(\mu/m_0)^{1/2}$ & $\kappa_{\rm lat}$ \\
\midrule
$\mathrm{AgBr}$ & $\mathrm{F\overline{4}3m}$ & 1 & 4 & 0.84 & 3.61 & 24.47 & 1.89 \\
$\mathrm{AgI}$ & $\mathrm{F\overline{4}3m}$ & 1 & 4 & 1.06 & 3.33 & 25.40 & 4.93 \\
$\mathrm{AgI}$ & $\mathrm{F\overline{4}3m}$ & 1 & 4 & 1.07 & 3.29 & 25.12 & 4.59 \\
$\mathrm{CuBr}$ & $\mathrm{F\overline{4}3m}$ & 1 & 4 & 0.82 & 4.58 & 27.27 & 11.38 \\
$\mathrm{CuCl}$ & $\mathrm{F\overline{4}3m}$ & 1 & 4 & 0.63 & 5.08 & 24.22 & 3.63 \\
$\mathrm{AgBr}$ & $\mathrm{Fm\overline{3}m}$ & 1 & 6 & 1.06 & 2.83 & 19.18 & 0.40 \\
$\mathrm{AgCl}$ & $\mathrm{Fm\overline{3}m}$ & 1 & 6 & 0.83 & 3.93 & 20.29 & 0.52 \\
$\mathrm{AgI}$ & $\mathrm{Fm\overline{3}m}$ & 1 & 6 & 1.46 & 2.23 & 17.06 & 0.34 \\
$\mathrm{CsBr}$ & $\mathrm{Fm\overline{3}m}$ & 1 & 6 & 0.57 & 2.40 & 16.94 & 0.88 \\
$\mathrm{CsCl}$ & $\mathrm{Fm\overline{3}m}$ & 1 & 6 & 0.50 & 3.31 & 17.51 & 1.36 \\
$\mathrm{CsF}$ & $\mathrm{Fm\overline{3}m}$ & 1 & 6 & 0.36 & 4.53 & 18.48 & 1.82 \\
$\mathrm{CsH}$ & $\mathrm{Fm\overline{3}m}$ & 1 & 6 & 0.50 & 13.41 & 13.41 & 2.36 \\
$\mathrm{CsI}$ & $\mathrm{Fm\overline{3}m}$ & 1 & 6 & 0.69 & 1.96 & 15.78 & 1.36 \\
$\mathrm{KBr}$ & $\mathrm{Fm\overline{3}m}$ & 1 & 6 & 0.46 & 3.51 & 17.98 & 2.52 \\
$\mathrm{KCl}$ & $\mathrm{Fm\overline{3}m}$ & 1 & 6 & 0.37 & 4.74 & 20.46 & 6.46 \\
$\mathrm{KF}$ & $\mathrm{Fm\overline{3}m}$ & 1 & 6 & 0.25 & 6.92 & 24.73 & 7.18 \\
$\mathrm{KH}$ & $\mathrm{Fm\overline{3}m}$ & 1 & 6 & 0.30 & 15.55 & 15.42 & 10.92 \\
$\mathrm{KI}$ & $\mathrm{Fm\overline{3}m}$ & 1 & 6 & 0.56 & 3.02 & 16.51 & 1.94 \\
$\mathrm{LiBr}$ & $\mathrm{Fm\overline{3}m}$ & 1 & 6 & 0.44 & 5.33 & 13.46 & 2.60 \\
$\mathrm{LiCl}$ & $\mathrm{Fm\overline{3}m}$ & 1 & 6 & 0.33 & 6.21 & 14.97 & 4.45 \\
$\mathrm{LiF}$ & $\mathrm{Fm\overline{3}m}$ & 1 & 6 & 0.16 & 9.61 & 21.66 & 16.54 \\
$\mathrm{LiH}$ & $\mathrm{Fm\overline{3}m}$ & 1 & 6 & 0.20 & 18.95 & 17.78 & 26.72 \\
$\mathrm{LiI}$ & $\mathrm{Fm\overline{3}m}$ & 1 & 6 & 0.64 & 4.84 & 12.42 & 1.70 \\
$\mathrm{NaBr}$ & $\mathrm{Fm\overline{3}m}$ & 1 & 6 & 0.41 & 4.38 & 18.51 & 2.85 \\
$\mathrm{NaCl}$ & $\mathrm{Fm\overline{3}m}$ & 1 & 6 & 0.31 & 5.92 & 22.11 & 6.49 \\
\bottomrule
\end{tabular*}
\end{table}
\addtocounter{table}{-1}
\begin{table}[p]
\centering
\caption{Materials selected for Fig.~\ref{fig_bond}(b) and Fig.~\ref{fig_group} (continued).}
\small
\setlength{\tabcolsep}{5pt}
\begin{tabular*}{\textwidth}{@{\extracolsep{\fill}}llrrrrrr}
\toprule
Formula & Space group & $V$ & $N_{\rm nn}$ & $\Delta\mathcal{G}$ & $\omega_{\rm TO}$ & $\omega_{\rm TO}(\mu/m_0)^{1/2}$ & $\kappa_{\rm lat}$ \\
\midrule
$\mathrm{NaF}$ & $\mathrm{Fm\overline{3}m}$ & 1 & 6 & 0.17 & 9.42 & 30.39 & 22.41 \\
$\mathrm{NaH}$ & $\mathrm{Fm\overline{3}m}$ & 1 & 6 & 0.22 & 19.05 & 18.72 & 18.67 \\
$\mathrm{NaI}$ & $\mathrm{Fm\overline{3}m}$ & 1 & 6 & 0.56 & 3.96 & 17.48 & 1.70 \\
$\mathrm{RbBr}$ & $\mathrm{Fm\overline{3}m}$ & 1 & 6 & 0.50 & 2.80 & 18.00 & 2.83 \\
$\mathrm{RbCl}$ & $\mathrm{Fm\overline{3}m}$ & 1 & 6 & 0.41 & 3.81 & 19.10 & 2.38 \\
$\mathrm{RbF}$ & $\mathrm{Fm\overline{3}m}$ & 1 & 6 & 0.29 & 5.73 & 22.58 & 3.11 \\
$\mathrm{RbH}$ & $\mathrm{Fm\overline{3}m}$ & 1 & 6 & 0.37 & 14.37 & 14.35 & 6.10 \\
$\mathrm{RbI}$ & $\mathrm{Fm\overline{3}m}$ & 1 & 6 & 0.60 & 2.31 & 16.52 & 1.53 \\
$\mathrm{TlBr}$ & $\mathrm{Fm\overline{3}m}$ & 1 & 6 & 1.20 & 2.00 & 15.12 & 1.50 \\
$\mathrm{TlCl}$ & $\mathrm{Fm\overline{3}m}$ & 1 & 6 & 0.93 & 2.55 & 14.00 & 0.88 \\
$\mathrm{TlI}$ & $\mathrm{Fm\overline{3}m}$ & 1 & 6 & 1.54 & 1.83 & 16.22 & 1.32 \\
$\mathrm{CsBr}$ & $\mathrm{Pm\overline{3}m}$ & 1 & 8 & 0.60 & 2.29 & 16.20 & 0.34 \\
$\mathrm{CsCl}$ & $\mathrm{Pm\overline{3}m}$ & 1 & 8 & 0.50 & 3.18 & 16.84 & 0.73 \\
$\mathrm{CsH}$ & $\mathrm{Pm\overline{3}m}$ & 1 & 8 & 0.53 & 11.97 & 11.97 & 3.74 \\
$\mathrm{CsI}$ & $\mathrm{Pm\overline{3}m}$ & 1 & 8 & 0.69 & 1.82 & 14.64 & 1.50 \\
$\mathrm{KBr}$ & $\mathrm{Pm\overline{3}m}$ & 1 & 8 & 0.46 & 2.75 & 14.10 & 1.32 \\
$\mathrm{KCl}$ & $\mathrm{Pm\overline{3}m}$ & 1 & 8 & 0.38 & 3.93 & 16.96 & 2.63 \\
$\mathrm{KF}$ & $\mathrm{Pm\overline{3}m}$ & 1 & 8 & 0.25 & 6.01 & 21.47 & 3.90 \\
$\mathrm{KI}$ & $\mathrm{Pm\overline{3}m}$ & 1 & 8 & 0.60 & 2.41 & 13.18 & 0.68 \\
$\mathrm{RbBr}$ & $\mathrm{Pm\overline{3}m}$ & 1 & 8 & 0.50 & 2.44 & 15.67 & 1.87 \\
$\mathrm{RbCl}$ & $\mathrm{Pm\overline{3}m}$ & 1 & 8 & 0.42 & 3.47 & 17.38 & 1.28 \\
$\mathrm{RbF}$ & $\mathrm{Pm\overline{3}m}$ & 1 & 8 & 0.29 & 5.13 & 20.23 & 2.41 \\
$\mathrm{RbI}$ & $\mathrm{Pm\overline{3}m}$ & 1 & 8 & 0.64 & 1.94 & 13.85 & 0.75 \\
$\mathrm{TlBr}$ & $\mathrm{Pm\overline{3}m}$ & 1 & 8 & 1.21 & 1.51 & 11.46 & 0.36 \\
$\mathrm{TlCl}$ & $\mathrm{Pm\overline{3}m}$ & 1 & 8 & 0.97 & 1.76 & 9.69 & 0.63 \\
\bottomrule
\end{tabular*}
\end{table}
\addtocounter{table}{-1}
\begin{table}[p]
\centering
\caption{Materials selected for Fig.~\ref{fig_bond}(b) and Fig.~\ref{fig_group} (continued).}
\small
\setlength{\tabcolsep}{5pt}
\begin{tabular*}{\textwidth}{@{\extracolsep{\fill}}llrrrrrr}
\toprule
Formula & Space group & $V$ & $N_{\rm nn}$ & $\Delta\mathcal{G}$ & $\omega_{\rm TO}$ & $\omega_{\rm TO}(\mu/m_0)^{1/2}$ & $\kappa_{\rm lat}$ \\
\midrule
$\mathrm{TlI}$ & $\mathrm{Pm\overline{3}m}$ & 1 & 8 & 1.47 & 1.06 & 9.34 & 0.21 \\
$\mathrm{BeO}$ & $\mathrm{F\overline{4}3m}$ & 2 & 4 & 0.27 & 22.42 & 53.84 & 338.81 \\
$\mathrm{BeS}$ & $\mathrm{F\overline{4}3m}$ & 2 & 4 & 0.55 & 17.39 & 46.13 & 176.24 \\
$\mathrm{BeSe}$ & $\mathrm{F\overline{4}3m}$ & 2 & 4 & 0.68 & 15.15 & 43.10 & 109.50 \\
$\mathrm{BeTe}$ & $\mathrm{F\overline{4}3m}$ & 2 & 4 & 0.91 & 13.99 & 40.60 & 88.47 \\
$\mathrm{CaSe}$ & $\mathrm{F\overline{4}3m}$ & 2 & 4 & 0.97 & 6.03 & 31.07 & 6.73 \\
$\mathrm{CdS}$ & $\mathrm{F\overline{4}3m}$ & 2 & 4 & 0.78 & 7.21 & 36.03 & 19.68 \\
$\mathrm{CdSe}$ & $\mathrm{F\overline{4}3m}$ & 2 & 4 & 1.40 & 5.14 & 34.98 & 22.57 \\
$\mathrm{CdTe}$ & $\mathrm{F\overline{4}3m}$ & 2 & 4 & 1.70 & 4.36 & 33.71 & 9.90 \\
$\mathrm{MgS}$ & $\mathrm{F\overline{4}3m}$ & 2 & 4 & 0.70 & 10.04 & 37.31 & 29.05 \\
$\mathrm{MgSe}$ & $\mathrm{F\overline{4}3m}$ & 2 & 4 & 0.86 & 8.12 & 35.00 & 20.65 \\
$\mathrm{MgTe}$ & $\mathrm{F\overline{4}3m}$ & 2 & 4 & 1.10 & 7.10 & 32.07 & 13.22 \\
$\mathrm{ZnO}$ & $\mathrm{F\overline{4}3m}$ & 2 & 4 & 0.63 & 12.29 & 44.08 & 58.22 \\
$\mathrm{ZnS}$ & $\mathrm{F\overline{4}3m}$ & 2 & 4 & 0.92 & 8.50 & 39.43 & 51.71 \\
$\mathrm{ZnSe}$ & $\mathrm{F\overline{4}3m}$ & 2 & 4 & 1.14 & 6.40 & 38.30 & 21.22 \\
$\mathrm{ZnTe}$ & $\mathrm{F\overline{4}3m}$ & 2 & 4 & 1.44 & 5.35 & 35.21 & 52.79 \\
$\mathrm{BaO}$ & $\mathrm{Fm\overline{3}m}$ & 2 & 6 & 1.00 & 3.49 & 13.22 & 3.18 \\
$\mathrm{BaS}$ & $\mathrm{Fm\overline{3}m}$ & 2 & 6 & 1.28 & 4.54 & 23.15 & 7.28 \\
$\mathrm{BaSe}$ & $\mathrm{Fm\overline{3}m}$ & 2 & 6 & 1.44 & 3.32 & 23.49 & 9.41 \\
$\mathrm{BaTe}$ & $\mathrm{Fm\overline{3}m}$ & 2 & 6 & 1.63 & 2.89 & 23.54 & 5.42 \\
$\mathrm{BeO}$ & $\mathrm{Fm\overline{3}m}$ & 2 & 6 & 0.31 & 6.62 & 15.90 & 13.02 \\
$\mathrm{CaO}$ & $\mathrm{Fm\overline{3}m}$ & 2 & 6 & 0.67 & 8.92 & 30.15 & 26.47 \\
$\mathrm{CaS}$ & $\mathrm{Fm\overline{3}m}$ & 2 & 6 & 0.94 & 7.07 & 29.86 & 27.45 \\
$\mathrm{CaSe}$ & $\mathrm{Fm\overline{3}m}$ & 2 & 6 & 1.09 & 5.49 & 28.32 & 15.35 \\
$\mathrm{CaTe}$ & $\mathrm{Fm\overline{3}m}$ & 2 & 6 & 1.31 & 4.90 & 27.07 & 10.63 \\
\bottomrule
\end{tabular*}
\end{table}
\addtocounter{table}{-1}
\begin{table}[p]
\centering
\caption{Materials selected for Fig.~\ref{fig_bond}(b) and Fig.~\ref{fig_group} (continued).}
\small
\setlength{\tabcolsep}{5pt}
\begin{tabular*}{\textwidth}{@{\extracolsep{\fill}}llrrrrrr}
\toprule
Formula & Space group & $V$ & $N_{\rm nn}$ & $\Delta\mathcal{G}$ & $\omega_{\rm TO}$ & $\omega_{\rm TO}(\mu/m_0)^{1/2}$ & $\kappa_{\rm lat}$ \\
\midrule
$\mathrm{GeTe}$ & $\mathrm{R3m}$ & 2 & 6 & 4.37 & 2.44 & 16.62 & 6.47 \\
$\mathrm{MgO}$ & $\mathrm{Fm\overline{3}m}$ & 2 & 6 & 0.40 & 12.64 & 39.25 & 56.47 \\
$\mathrm{MgO}$ & $\mathrm{P\overline{6}m2}$ & 2 & 6 & 0.44 & 10.74 & 33.37 & 17.70 \\
$\mathrm{MgS}$ & $\mathrm{Fm\overline{3}m}$ & 2 & 6 & 0.80 & 7.49 & 27.86 & 21.04 \\
$\mathrm{MgSe}$ & $\mathrm{Fm\overline{3}m}$ & 2 & 6 & 1.05 & 5.83 & 25.15 & 9.92 \\
$\mathrm{MgTe}$ & $\mathrm{Fm\overline{3}m}$ & 2 & 6 & 1.52 & 4.74 & 21.43 & 4.01 \\
$\mathrm{PbS}$ & $\mathrm{Fm\overline{3}m}$ & 2 & 6 & 2.79 & 2.70 & 14.24 & 3.40 \\
$\mathrm{PbSe}$ & $\mathrm{Fm\overline{3}m}$ & 2 & 6 & 3.26 & 1.89 & 14.26 & 4.26 \\
$\mathrm{PbTe}$ & $\mathrm{Fm\overline{3}m}$ & 2 & 6 & 4.11 & 1.51 & 13.45 & 3.01 \\
$\mathrm{SnTe}$ & $\mathrm{Fm\overline{3}m}$ & 2 & 6 & 4.98 & 1.44 & 11.27 & 5.99 \\
$\mathrm{SrO}$ & $\mathrm{Fm\overline{3}m}$ & 2 & 6 & 0.78 & 6.73 & 24.74 & 11.83 \\
$\mathrm{SrS}$ & $\mathrm{Fm\overline{3}m}$ & 2 & 6 & 1.04 & 5.70 & 27.62 & 11.61 \\
$\mathrm{SrSe}$ & $\mathrm{Fm\overline{3}m}$ & 2 & 6 & 1.16 & 4.16 & 26.79 & 15.60 \\
$\mathrm{SrTe}$ & $\mathrm{Fm\overline{3}m}$ & 2 & 6 & 1.39 & 3.61 & 26.03 & 10.19 \\
$\mathrm{YbS}$ & $\mathrm{Fm\overline{3}m}$ & 2 & 6 & 0.97 & 6.19 & 32.19 & 26.26 \\
$\mathrm{YbTe}$ & $\mathrm{Fm\overline{3}m}$ & 2 & 6 & 1.35 & 3.40 & 29.15 & 9.32 \\
$\mathrm{ZnO}$ & $\mathrm{Fm\overline{3}m}$ & 2 & 6 & 0.64 & 8.03 & 28.80 & 9.32 \\
$\mathrm{BaTe}$ & $\mathrm{Pm\overline{3}m}$ & 2 & 8 & 1.80 & 2.12 & 17.23 & 1.26 \\
$\mathrm{AlAs}$ & $\mathrm{F\overline{4}3m}$ & 3 & 4 & 1.34 & 10.49 & 46.70 & 93.48 \\
$\mathrm{AlN}$ & $\mathrm{F\overline{4}3m}$ & 3 & 4 & 0.61 & 19.82 & 60.18 & 224.30 \\
$\mathrm{AlP}$ & $\mathrm{F\overline{4}3m}$ & 3 & 4 & 1.13 & 12.90 & 49.00 & 80.39 \\
$\mathrm{AlSb}$ & $\mathrm{F\overline{4}3m}$ & 3 & 4 & 1.69 & 9.24 & 43.42 & 98.43 \\
$\mathrm{BAs}$ & $\mathrm{F\overline{4}3m}$ & 3 & 4 & 0.50 & 20.61 & 63.35 & 1343.59 \\
$\mathrm{BN}$ & $\mathrm{F\overline{4}3m}$ & 3 & 4 & 0.27 & 31.87 & 78.72 & 678.80 \\
$\mathrm{BP}$ & $\mathrm{F\overline{4}3m}$ & 3 & 4 & 0.41 & 24.08 & 68.16 & 520.58 \\
\bottomrule
\end{tabular*}
\end{table}
\addtocounter{table}{-1}
\begin{table}[p]
\centering
\caption{Materials selected for Fig.~\ref{fig_bond}(b) and Fig.~\ref{fig_group} (continued).}
\small
\setlength{\tabcolsep}{5pt}
\begin{tabular*}{\textwidth}{@{\extracolsep{\fill}}llrrrrrr}
\toprule
Formula & Space group & $V$ & $N_{\rm nn}$ & $\Delta\mathcal{G}$ & $\omega_{\rm TO}$ & $\omega_{\rm TO}(\mu/m_0)^{1/2}$ & $\kappa_{\rm lat}$ \\
\midrule
$\mathrm{GaAs}$ & $\mathrm{F\overline{4}3m}$ & 3 & 4 & 1.75 & 7.85 & 47.16 & 39.80 \\
$\mathrm{GaN}$ & $\mathrm{P6_{3}mc}$ & 3 & 4 & 0.80 & 4.05 & 13.83 & 208.48 \\
$\mathrm{GaN}$ & $\mathrm{F\overline{4}3m}$ & 3 & 4 & 0.83 & 16.30 & 55.67 & 201.42 \\
$\mathrm{GaP}$ & $\mathrm{F\overline{4}3m}$ & 3 & 4 & 1.46 & 10.65 & 49.34 & 115.54 \\
$\mathrm{PrN}$ & $\mathrm{F\overline{4}3m}$ & 3 & 4 & 1.99 & 12.33 & 44.00 & 42.23 \\
$\mathrm{ScN}$ & $\mathrm{F\overline{4}3m}$ & 3 & 4 & 1.18 & 15.53 & 50.75 & 70.42 \\
$\mathrm{ScP}$ & $\mathrm{F\overline{4}3m}$ & 3 & 4 & 1.46 & 9.65 & 41.33 & 10.17 \\
$\mathrm{AlN}$ & $\mathrm{Fm\overline{3}m}$ & 3 & 6 & 0.58 & 14.61 & 44.37 & 75.28 \\
$\mathrm{GaN}$ & $\mathrm{Fm\overline{3}m}$ & 3 & 6 & 1.00 & 7.91 & 27.02 & 7.06 \\
$\mathrm{C}$ & $\mathrm{Fd\overline{3}m}$ & 4 & 4 & -0.09 & 39.22 & 96.10 & 1799.67 \\
$\mathrm{Ge}$ & $\mathrm{Fd\overline{3}m}$ & 4 & 4 & 2.04 & 8.24 & 49.63 & 52.00 \\
$\mathrm{GeC}$ & $\mathrm{F\overline{4}3m}$ & 4 & 4 & 0.99 & 19.69 & 63.20 & 260.11 \\
$\mathrm{Si}$ & $\mathrm{Fd\overline{3}m}$ & 4 & 4 & 1.08 & 15.08 & 56.51 & 121.63 \\
$\mathrm{SiC}$ & $\mathrm{F\overline{4}3m}$ & 4 & 4 & 0.69 & 23.47 & 68.09 & 446.17 \\
$\mathrm{SiGe}$ & $\mathrm{F\overline{4}3m}$ & 4 & 4 & 1.30 & 12.12 & 54.55 & 120.43 \\
$\mathrm{SiSn}$ & $\mathrm{F\overline{4}3m}$ & 4 & 4 & 2.11 & 10.26 & 48.89 & 75.74 \\
$\mathrm{SnC}$ & $\mathrm{F\overline{4}3m}$ & 4 & 4 & 1.55 & 16.57 & 54.74 & 160.47 \\
\bottomrule
\end{tabular*}
\end{table}